%% file: main.tex
\documentclass[sigconf, nonacm]{acmart}
\usepackage{numprint} 
\usepackage{array}
\usepackage{listings}
\usepackage{xspace}
\usepackage[table, dvipsnames]{xcolor}
\usepackage{cleveref}
\usepackage{multirow}
\usepackage{snaptodo}

\usepackage{subcaption}
\usepackage{makecell}
\usepackage{fancyhdr}

\usepackage{booktabs}
\usepackage{array}
\usepackage{tikz}
\usepackage{threeparttable}
\usepackage{amsmath}
\usepackage[usestackEOL]{stackengine}
\usepackage{pgfplots, pgfplotstable}
\pgfplotsset{compat=1.18}

\usepackage{enumitem}

\usepackage{rotating} 
\usepackage{graphicx} 

\usepackage[most]{tcolorbox}
\usepackage{xcolor}

\usepackage{siunitx}

\usepackage[framemethod=default]{mdframed}

\newmdenv[
  backgroundcolor=blue!7,
  linewidth=0pt,
  leftmargin=0pt,
  rightmargin=0pt,
  innerleftmargin=3pt,
  innerrightmargin=3pt,
  innertopmargin=3pt,
  innerbottommargin=3pt,
  skipabove=0.6\baselineskip,
  skipbelow=0.6\baselineskip,
  splittopskip=\baselineskip,
  splitbottomskip=\baselineskip,
]{reviewbox}

\newcommand{\SemBenchHeader}{%
  \toprule
    & \multicolumn{3}{c}{\textbf{LOTUS}} & \multicolumn{3}{c}{\textbf{Palimpzest}}
    & \multicolumn{3}{c}{\textbf{ThalamusDB}} & \multicolumn{3}{c}{\textbf{BigQuery}} \\
  \cmidrule(lr){2-4} \cmidrule(lr){5-7} \cmidrule(lr){8-10} \cmidrule(lr){11-13}
    & \textbf{Cost} & \textbf{Quality} & \textbf{Latency}
    & \textbf{Cost} & \textbf{Quality} & \textbf{Latency}
    & \textbf{Cost} & \textbf{Quality} & \textbf{Latency}
    & \textbf{Cost} & \textbf{Quality} & \textbf{Latency} \\
  \midrule
}

\newcommand{\ScenarioRow}[1]{%
  \addlinespace[0.4ex]
  \multicolumn{13}{c}{\textbf{#1}}\\
  \addlinespace[0.4ex]
}

\newcommand{\ContinuedNote}{%
  \multicolumn{13}{r}{\small\itshape Table~\ref{tab:experimental_results_all} continued from previous page.}\\[-0.3ex]
}

\usepackage{amssymb}
\usepackage{pifont}
\newcommand{\xmark}{\ding{55}}%

\newcommand\vldbdoi{XX.XX/XXX.XX}
\newcommand\vldbpages{XXX-XXX}
\newcommand\vldbvolume{14}
\newcommand\vldbissue{1}
\newcommand\vldbyear{2020}
\newcommand\vldbauthors{\authors}
\newcommand\vldbtitle{\shorttitle} 
\newcommand\vldbavailabilityurl{https://sembench.org}
\newcommand\vldbpagestyle{plain} 

\def\arraystretchresulttable{0.8}

\begin{document}



\title{SemBench: A Benchmark for Semantic Query Processing Engines [Experiment, Analysis \& Benchmark]}


\author{Jiale Lao}
\orcid{0009-0003-1144-5152}
\affiliation{%
  \institution{Cornell University}
}
\email{jiale@cs.cornell.edu}

\author{Andreas Zimmerer}
\orcid{0000-0002-4158-5805}
\affiliation{%
  \institution{University of Technology Nuremberg}
}
\email{andreas.zimmerer@utn.de}

\author{Olga Ovcharenko}
\orcid{0009-0003-3676-482X}
\affiliation{%
  \institution{BIFOLD \& TU Berlin}
}
\email{ovcharenko@tu-berlin.de}

\author{Tianji Cong}
\orcid{0000-0002-7793-7059}
\affiliation{%
  \institution{University of Michigan}
}
\email{congtj@umich.edu}

\author{Matthew Russo}
\orcid{0009-0005-9685-3976}
\affiliation{%
  \institution{MIT CSAIL}
}
\email{mdrusso@mit.edu}

\author{Gerardo Vitagliano}
\orcid{0000-0001-7782-2596}
\affiliation{%
  \institution{MIT CSAIL}
}
\email{gerarvit@mit.edu}

\author{Michael Cochez}
\orcid{0000-0001-5726-4638}
\affiliation{%
  \institution{Vrije Universiteit Amsterdam}
}
\email{m.cochez@vu.nl}

\author{Fatma \"Ozcan}
\orcid{}
\affiliation{%
  \institution{Google}
}
\email{fozcan@google.com}

\author{Gautam Gupta}
\orcid{}
\affiliation{%
  \institution{Google}
}
\email{gautamguptag@google.com}

\author{Thibaud Hottelier}
\orcid{}
\affiliation{%
  \institution{Google}
}
\email{tbh@google.com}


\author{H. V. Jagadish}
\orcid{}
\affiliation{%
  \institution{University of Michigan}
}
\email{jag@umich.edu}

\author{Kris Kissel}
\orcid{}
\affiliation{%
  \institution{Google}
}
\email{kriskissel@google.com}

\author{Sebastian Schelter}
\orcid{0000-0003-4722-5840}
\affiliation{%
  \institution{BIFOLD \& TU Berlin}
}
\email{schelter@tu-berlin.de}

\author{Andreas Kipf}
\orcid{0000-0003-3463-0564}
\affiliation{%
  \institution{University of Technology Nuremberg}
}
\email{andreas.kipf@utn.de}

\author{Immanuel Trummer}
\orcid{0000-0002-1579-3221}
\affiliation{%
  \institution{Cornell University}
}
\email{itrummer@cornell.edu}



\begin{abstract}
We present a benchmark targeting a novel class of systems: semantic query processing engines. Those systems rely inherently on generative and reasoning  capabilities of state-of-the-art large language models (LLMs). They extend SQL with semantic operators, configured by natural language instructions, that are evaluated via LLMs and enable users to perform various operations on multimodal data.



Our benchmark introduces diversity across three key dimensions: scenarios, modalities, and operators.
Included are scenarios ranging from movie review analysis to {car damage detection.}
Within these scenarios, we cover different data modalities, including images, audio, and text. 
Finally, the queries involve a diverse set of operators, including semantic filters, joins, mappings, ranking, and classification operators. 

We evaluated our benchmark on three academic systems (LOTUS, Palimpzest, and ThalamusDB) and one industrial system, Google BigQuery. Although these results reflect a snapshot of systems under continuous development, our study offers crucial insights into their current strengths and weaknesses, illuminating promising directions for future research.

\end{abstract}

\input{settings}

\maketitle

\pagestyle{\vldbpagestyle}
\begingroup\small\noindent\raggedright\textbf{PVLDB Reference Format:}\\
\vldbauthors. \vldbtitle. PVLDB, \vldbvolume(\vldbissue): \vldbpages, \vldbyear.\\
\href{https://doi.org/\vldbdoi}{doi:\vldbdoi}
\endgroup
\begingroup
\renewcommand\thefootnote{}\footnote{\noindent
This work is licensed under the Creative Commons BY-NC-ND 4.0 International License. Visit \url{https://creativecommons.org/licenses/by-nc-nd/4.0/} to view a copy of this license. For any use beyond those covered by this license, obtain permission by emailing \href{mailto:info@vldb.org}{info@vldb.org}. Copyright is held by the owner/author(s). Publication rights licensed to the VLDB Endowment. \\
\raggedright Proceedings of the VLDB Endowment, Vol. \vldbvolume, No. \vldbissue\ %
ISSN 2150-8097. \\
\href{https://doi.org/\vldbdoi}{doi:\vldbdoi} \\
}\addtocounter{footnote}{-1}\endgroup

\ifdefempty{\vldbavailabilityurl}{}{
\vspace{.3cm}
\begingroup\small\noindent\raggedright\textbf{PVLDB Artifact Availability:}\\
The source code, data, and/or other artifacts have been made available at \url{\vldbavailabilityurl}.
\endgroup
}

\input{sections/01_introduction}
\input{sections/02_background} 
\input{sections/03_principles}
\input{sections/04_scenarios}
\input{sections/06_experimental_results}
\input{sections/07_conclusions}


\begin{acks}
The idea of a benchmark for semantic query processing originated at the Dagstuhl Seminar 25182 on Challenges and Opportunities of Table Representation Learning. We thank the seminar organizers for bringing this group of people together.
\end{acks}


\nocite{*}

\bibliographystyle{ACM-Reference-Format}
\bibliography{literature}
\balance

\end{document}

%% file: settings.tex

\newcommand*{\eg}{e.g.,\@\xspace}
\newcommand*{\ie}{i.e.,\@\xspace}
\newcommand*{\cf}{cf.\@\xspace}
\newcommand*{\etal}{~et~al.\@\xspace}
\newcommand*{\dash}{\textemdash\@\xspace}
\newcommand*{\etc}{~etc.\@\xspace}

\renewcommand*{\s}{\,s\@\xspace}
\renewcommand*{\ms}{\,ms\@\xspace}
\renewcommand*{\GHz}{\,GHz\@\xspace}
\newcommand*{\B}{\,B\@\xspace}
\newcommand*{\KB}{\,KB\@\xspace}
\newcommand*{\MB}{\,MB\@\xspace}
\newcommand*{\GB}{\,GB\@\xspace}
\newcommand*{\TB}{\,TB\@\xspace}
\renewcommand*{\percent}{\,\%\@\xspace}
\newcommand*{\M}{\,M\@\xspace}

\clubpenalty = 10000
\widowpenalty = 10000

\newcommand\paragraphnospace[1]{\noindent{\bfseries#1\space}}
\newcolumntype{C}[1]{>{\centering\arraybackslash}p{#1}}

\newcommand{\watermarked}[2]{
    \begin{tikzpicture}
        \node[opacity=1.0] at (0,0) {#1};
        \node[rotate=45, color=red, font=\Huge, opacity=0.4] at (0,0) {#2};
    \end{tikzpicture}
}
\newcommand{\draftwatermarked}[1]{\watermarked{#1}{DRAFT}}

\definecolor{dkgreen}{rgb}{0,0.6,0}
\definecolor{gray}{rgb}{0.5,0.5,0.5}
\definecolor{mauve}{rgb}{0.58,0,0.82}
\lstset{language=SQL,
  basicstyle={\small\ttfamily},
  belowskip=3mm,
  breakatwhitespace=true,
  breaklines=true,
  classoffset=0,
  columns=flexible,
  commentstyle=\color{dkgreen},
  framexleftmargin=0.25em,
  frameshape={}{}{}{}, 
  keywordstyle=\color{blue},
  numbers=none, 
  numberstyle=\tiny\color{gray},
  showstringspaces=false,
  stringstyle=\color{mauve},
  tabsize=3,
  xleftmargin =1em,
  literate={\\\%}{\%}1, 
}

\newcommand*\circled[1]{\tikz[baseline=(char.base)]{
            \node[shape=circle,draw,fill=white,inner sep=2pt] (char) {#1};}}

\newcommand{\forGoogleReview}[1]{\textcolor{blue}{#1}}
\newcommand{\oo}[1]{\textcolor{pink}{#1}}

\usetikzlibrary{patterns, 
    fit,
    patterns.meta,
    backgrounds, 
    }

\tikzset{
    fancycellbase/.style={
            rounded corners=2pt,
            inner sep=2pt,
            text height=1.3ex,
            text depth=.0ex,
            text centered,
            font=\bfseries,
            text=black
    }}
    
\tikzset{
    fancycellhatchbase/.style={
            opacity=0.5,
            rounded corners=2pt,
            fit=(X),
            inner sep=-0.1pt,
    }}

\tikzset{
    fancycellhatch1/.style={
            fancycellhatchbase,
            opacity=0.7,
            pattern={
                Lines[angle=-45,
                distance={3pt}, 
                line width=1pt]},
            pattern color = white,
    }
}

\tikzset{
    fancycellhatch2/.style={
            fancycellhatchbase,
            opacity=0.7,
            pattern={
                Lines[angle=90,
                distance={3pt}, 
                line width=1pt]},
            pattern color = white,
    }
}

\tikzset{
    fancycellhatch3/.style={
            fancycellhatchbase,
            opacity=0.7,
            pattern={
                Dots[angle=90,
                distance={3pt}, 
                radius=0.8pt]},
            pattern color = white,
    }
}

\tikzset{
    fancycellhatch4/.style={
            fancycellhatchbase,
            opacity=0.7,
            pattern={
                Hatch[angle=45,
                distance={6pt}, 
                line width=0.7pt]},
            pattern color = white,
    }
}

\newcommand{\fancycellC}[5]{%
  \begingroup
    \definecolor{__tmpfill}{HTML}{#2}%
    \begin{tikzpicture}[baseline=(X.base)]
        \node[fancycellbase,
            minimum width=#1,
        ] (X) {#3};
        \begin{scope}[on background layer]
            \node[fancycellhatchbase, fill=__tmpfill!70]{};
            \node[#4] {}; 
        \end{scope}
    \end{tikzpicture}%
  \endgroup
}

\newcommand{\fancycell}[3][3.6em]{\fancycellC{#1}{#2}{#3}{fancycellhatch3}{ }}

\newcommand{\fancycellGreen}[1]{\fancycellC{3.8em}{1e7e34}{#1}{fancycellhatch1}{ $\checkmark$ }}
\newcommand{\fancycellYellow}[1]{\fancycellC{3.8em}{b8860b}{#1}{fancycellhatch2}{ $\blacktriangle$ }}
\newcommand{\fancycellGray}[1]{\fancycellC{3.8em}{6c757d}{#1}{fancycellhatch3}{ $\bullet$ }}
\newcommand{\fancyCellRed}[1]{\fancycellC{0em}{c81b4a}{ \xmark }{fancycellhatch4}{}}

\newcommand{\fancycellNormal}[1]{\fancycellC{3.8em}{dddddd}{#1}{fancycellhatchbase}{  }}

\newcommand{\fancycellnormal}[3][3.6em]{%
  \begingroup
    \begin{tikzpicture}[baseline=(X.base)]
        \node[fancycellbase,
            minimum width=#1,
        ] (X) {#3};
        \node[] (Y) at (X.east){};
        \begin{scope}[on background layer]
            \node[fancycellhatchbase, fill=#2]{};
        \end{scope}
    \end{tikzpicture}%
  \endgroup
}

%% file: sections/01_introduction.tex
\section{Introduction}
\label{s:introduction}


With the advent of LLMs and their inherent generative and reasoning power, a novel class of data processing systems has emerged in academia~\cite{patel2024lotus, liu2025palimpzest, jo2024thalamusdb, urban2024caesura, dorbani2025flockmtl} and is already experiencing widespread adoption in industry~\cite{fernandes2015bigquery,dageville2016snowflake}: semantic query processing engines (SQPEs).
SQPEs extend relational algebra by \textit{semantic operators} that perform tasks described in natural language on a multitude of data types, including standard SQL types as well as unstructured data types such as images or audio data (stored in the cells of database tables). Such operators are enabled by the latest generation of LLMs, able to process multimodal inputs in \textit{zero-shot mode}, i.e., processing novel tasks based on a task description alone (without requiring any task-specific training data). The following query can be processed by several of the recently released SQPEs.

\begin{example}
Consider a table {\tt Cars} with a column {\tt pic} containing pictures of cars (each table row contains one picture). The query below, formulated in semantic SQL, counts pictures of red cars:
\begin{lstlisting}[
  language=SQL,
  showspaces=false,
  basicstyle=\ttfamily,
  numbers=left,
  numberstyle=\tiny,
  commentstyle=\color{gray}
]
SELECT COUNT(*) 
FROM Cars 
WHERE AI.IF(pic, 'the picture shows a red car');
\end{lstlisting}
\end{example}

The query above uses a semantic operator ({\tt AI.IF}) to process images according to natural language instructions (``the picture shows a red car''). While we use BigQuery's syntax in this and the following examples, the query above can be processed (after slight rewriting) by many other SQPEs as well (e.g., LOTUS, Palimpzest, and ThalamusDB).

Semantic operators are evaluated using LLMs, changing query optimization in two significant ways. First, when using LLMs, per-byte processing costs are higher by many orders of magnitude compared to traditional, relational processing. Hence, as soon as LLMs are invoked during query evaluation, the focus in cost optimization shifts from optimizing relational processing to minimizing costs associated with LLM invocations. Second, LLMs results are inherently stochastic, and  do not always produce 100\% correct results. Hence, accuracy of results becomes another important metric to optimize. For SQPEs, the  optimization problem can be stated as ``minimize the latency and cost of LLMs, while maximizing accuracy''. 

Current benchmarks for analytical data processing (e.g., TPC-H and TPC-DS) do not contain queries with semantic operators.
Over many decades, these benchmarks have driven innovation in areas that contribute to efficient relational processing, e.g., relational operator implementations, query optimization, and table indexing. However, they are unsuitable for guiding the development of techniques that minimize the number of LLM calls, the number of input tokens, or the size of the LLMs used for specific tasks during query processing. However, in the context of SQPEs, such techniques are crucial for efficient processing.

We introduce a new benchmark, SemBench, that focuses on query processing with semantic operators. The goal is to foster innovation in areas that contribute to minimizing costs associated with LLM invocations, while maximizing accuracy. Our benchmark refers to an extended relational data model that is nowadays adopted by most SQPEs: data are stored in relational tables, but cells may contain images or audio files in addition to the usual SQL data types. Semantic operators can be applied to images, text, audio files, or combinations of the aforementioned types. For instance, our benchmark features queries with semantic joins that cross data modalities (e.g., a join connecting images with associated text).

Our benchmark encompasses multiple scenarios, each characterized by a scenario-specific database and a set of queries. To create benchmark data, we combine existing data sets from Kaggle that come with manual labels. For instance, to create a database for a wildlife monitoring scenario, we use a data set from Kaggle with animal sound recordings, labeled by the associated species, as well as a dataset with camera trap images, each image labeled with the animal species that appear in it. By combining both data sets and generating additional metadata (e.g., to represent a recording location), we obtain a database that enables complex queries. For instance, we count locations at which specific animal species co-occur, based on detections inferred from audio or visual data. To generate ground-truth results for such queries, we leverage the manual labels of the original data sets.

Our benchmark queries cover a wide range of semantic operators (as well as traditional relational operators). It contains queries that are supported by all or most of the current SQPEs, as well as queries containing semantic operators only available in a subset of evaluated systems. Our goal is to motivate SQPE developers to expand their scope in terms of supported queries, as well as to evaluate the impact of various query processing approaches on performance and result quality. 

Our benchmark evaluates systems according to multiple metrics. First, we evaluate result quality by comparison to the ground truth. Depending on the type of query, we use different metrics to assess the similarity of ground truth and generated results. For aggregation queries (i.e., queries containing SQL aggregates), we measure quality as the relative error. For retrieval queries, we measure quality via F1 score (based on whether or not generated result rows appear in the ground truth). Second, we evaluate the processing overheads using metrics such as execution time, monetary execution fees (paid for LLM invocations), {and memory usage. Finally, we evaluate and compare the scalability of different systems.}

We present experimental results for SQPEs from academia and industry. From academia, we evaluate LOTUS~\cite{patel2024lotus}, Palimpzest~\cite{liu2025palimpzest}, and ThalamusDB~\cite{jo2024thalamusdb}. 
From industry, we evaluate Google's BigQuery system~\cite{fernandes2015bigquery}, now offering support for semantic operators (``AI Functions''). All evaluated systems are currently undergoing rapid changes, and our results should be understood merely as snapshots. We publish an online website for our benchmark\footnote{\url{https://sembench.org}} that will be regularly updated as new results become available. We analyze our initial  experimental results to derive insights regarding the strengths and weaknesses of the evaluated systems, the impact of specific performance optimization techniques, as well as avenues for future research. In summary, our original scientific contributions in this paper are the following:

\begin{itemize}
    \item We introduce a new benchmark that focuses on the emerging class of semantic query processing systems. It features queries with semantic operators on multimodal data. The benchmark contains 5 scenarios and 55 queries, covering analysis across three modalities: text, image, and audio. 
    \item We present an initial experimental study of the benchmark for a variety of semantic query processing engines, including systems from industry as well as from academia.
    \item We analyze these experimental results, link performance differences to specific query properties, and study the impact of different performance optimization techniques.
\end{itemize}

The remainder of this paper is organized as follows. In Section~\ref{sec:background}, we provide background and discuss related work. In Section~\ref{sec:principles}, we discuss the principles underlying our benchmark design. Next, in Section~\ref{sec:scenarios}, we describe the specific benchmark scenarios we created in detail. 
Section~\ref{sec:experiments} reports experimental results for different benchmarks and systems. Finally, Section~\ref{sec:conclusions} summarizes our experimental results and discusses future directions.




%% file: sections/02_background.tex
\section{Background}
\label{sec:background}

We discuss developments that led towards semantic query processing engines. Also, we discuss prior benchmarks and the differences.


\subsection{Crowdsourced Data Processing}

The idea of semantic operators is not new. The vision of expanding SQL with operators configured in natural language originated in the early 2010's in the area of crowdsourced database systems. Systems such as CrowdDB~\cite{Franklin2011}, Deco~\cite{Parameswaran2012}, or Qurk~\cite{Marcus2011} use human crowdworkers, hired on platform such as Amazon Mechanical Turk~\cite{AMTlink}, to perform tasks on multimodal data according to user instructions. Many of those systems support SQL-style query interfaces that enable users to include instructions (automatically conveyed to crowd workers during processing) as a part of their queries. At the time, only human workers were flexible enough to perform diverse tasks with various types of data according to instructions formulated in natural language. The high cost and latency that comes with using crowdworkers has inspired a large body of research, aimed at making crowdsourced database systems more efficient (e.g., via specialized operator implementations~\cite{Marcus2011b, Marcus2012, DasSarma2014} or query optimization variants~\cite{Fan2015, Parameswaran2012, Li2017}). Despite those advances, processing data with crowdworkers remains expensive and, due to the limited number of crowdworkers, the scalability of the approach is limited.

\subsection{Large Language Models}

Large language models (LLMs) have enabled stunning advances in multiple areas of computer science over the past years. Those advances have been fueled by two key ideas: the Transformer architecture~\cite{Vaswani2017}, making it easier to scale models up to large parameter counts, and the idea of transfer learning~\cite{ruder2019transfer}, i.e., training models on generic tasks for which large amounts of (unlabeled) training data are readily available, hoping for a knowledge transfer to other tasks where training data is sparse. For instance, current LLMs are often trained on the task of predicting the next word (or, more precisely, token, the atomic unit at which language models represent text) in arbitrary web text~\cite{GPT2}. To perform well on that task, LLMs must develop capabilities akin to a rudimentary level of natural language understanding, as well as commonsense knowledge. Those skills are useful for many other tasks as well. While early-stage LLMs required fine-tuning for specific tasks, recent LLMs trained on large-scale generic corpora can address new tasks~\cite{Brown2020}. They can do so based on a natural language description alone (zero-shot learning), and their performance can further improve when provided with a few examples (few-shot learning) or when leveraging their reasoning ability. Those capabilities have enabled a new generation of systems, discussed next.

\subsection{Semantic Query Processing Engines}

Crowdsourced database systems introduced semantic operators. However, their scalability is limited due to their reliance on crowdworkers. Over the past two years, a new generation of database systems have appeared that support semantic operators but rely on LLMs to evaluate them. These systems include several research prototypes from academia~\cite{patel2024lotus, liu2025palimpzest, jo2024thalamusdb, dorbani2025flockmtl, urban2024caesura}, as well as several industry systems by major Cloud providers, in particular Google's BigQuery~\cite{fernandes2015bigquery} and Snowflake's platform~\cite{dageville2016snowflake}, now supporting semantic operators as well. 

\subsection{Related Benchmarks}

Our benchmark focuses on analytical data processing. In that, it relates to other popular benchmarks in the database community, for instance, TPC-H~\cite{TPC2013}, TPC-DS~\cite{tpcds}, and the Join Order Benchmark~\cite{Gubichev2015}. However, all of the aforementioned benchmarks focus on purely relational data processing. In that context, overheads due to data movements and relational operator evaluations are dominant. Instead, SemBench focuses on queries with semantic operators. For such queries, the overheads due to LLM calls for semantic operator evaluations are dominant. Hence, our benchmark motivates a very different set of techniques to optimize performance.

SemBench focuses on multimodal data processing via LLMs. In that, it relates to prior work proposing benchmarks for multimodal question answering~\cite{Bae2023, Li2024b, Liu2025}. However, prior benchmarks in this domain typically aim at evaluating multimodal models, focusing on output quality rather than cost and execution time. Instead, SemBench aims at evaluating systems for semantic query processing, measuring processing overheads, along with quality. It considers complex queries mixing semantic operators (evaluated via language models) with traditional SQL operators.







%% file: sections/03_principles.tex
\section{Benchmark Design Principles}\label{sec:principles}

{SemBench is designed to use data modalities, operators, and scenarios that are already adopted by many systems in academia and industry. Each query can be processed by at least two semantic query processing engines~\cite{patel2024lotus, liu2025palimpzest, jo2024thalamusdb, fernandes2015bigquery, dageville2016snowflake, dorbani2025flockmtl, urban2024caesura}. While some systems currently support only a subset due to missing operators or data modalities, they are under active development. The selected scenarios are diverse, covering common evaluation settings used in academia, and are inspired by workloads observed in industry.
}


\subsection{Data}

For our benchmark, we assume a relational data model with an extended type system. Beyond the SQL standard data types, columns may also contain images or audio files. For our benchmark, we store the path to the associated (image or audio) files in the corresponding column. All evaluated SQPEs support this data format.

Our benchmark data is based on manually annotated data sets, created for tasks such as image, text, or audio data classification. We combine such data sets (obtained primarily from the Kaggle platform), in some cases enriched by randomly generated data, to generate databases for our benchmark scenarios. By leveraging ground truth labels that come with the original datasets, we can obtain ground truth results for our benchmark queries.

Most of our benchmark databases contain data of multiple modalities (e.g., images and text). However, we also include scenarios that focus on one single modality, in particular text, in addition to tabular data. This enables us to evaluate SQPEs that support only a subset of data types.


\subsection{Queries}

We generate at least ten queries for each benchmark scenario. Each of our queries contains semantic operators, possibly mixed with standard SQL operations. For all of our queries, processing overheads due to semantic operators dominate total computation costs. This reflects the focus of our benchmark on techniques that reduce the overheads of semantic query processing.

Our queries cover a wide range of semantic operators. Table~\ref{tab:semantic-operators} summarizes the semantic operators supported by SemBench. These operators include semantic filters, which select multimodal data based on natural language instructions; semantic joins, which identify item pairs that satisfy specified matching conditions; semantic maps, which transform data according to natural language descriptions; semantic ranks, which order items based on text-described criteria; and semantic classify, which assigns items to predefined categories. The set of supported semantic operators differs across SQPEs. By analyzing available engines, we identified operators that are supported by most current SQPEs, as well as operators for which support is sparse. Our goal in creating the benchmark queries was to include queries that can be evaluated by the majority of SQPEs, as well as some queries that require more ``exotic'' operators (possibly motivating the inclusion of such operators in future system versions). Table~\ref{tab:operators} defines all semantic operators used in our benchmark, along with examples.


\begin{table*}[t]
  \centering
  \caption{Summary of key semantic operators covered by SemBench. $T$ is a relation, $X,Y$ are tuple types, $t$ is a specific tuple, $l$ is a natural language expression, and $M$ is a model processing tuples based on $l$. 
  We adopt the syntax from~\cite{patel2024lotus} for its simplicity and clarity, and we redefine and extend several operators for better generalization.\label{tab:operators}
  }
  \label{tab:semantic-operators}
  \renewcommand{\arraystretch}{1.2}
  \setlength{\tabcolsep}{2pt}
  \begin{tabular}{p{0.25\linewidth} p{0.32\linewidth} p{0.38\linewidth}}
    \toprule
    \textbf{Operator} & \textbf{Definition} & \textbf{Example} \\
    \midrule
    \texttt{sem\_filter($l:X\!\to\! Bool$)} &
      $\{\,t \in T \mid M(l)(t)=1\,\}$ &
      Select patients with sick lung based on X-ray images \\[0.6ex]
    \texttt{sem\_join($l:(X,Y)\!\to\! Bool$)} &
      $\{(t_i,t_j) \mid M(l)(t_i,t_j)=1,\,t_i\in T_1,\,t_j\in T_2\}$ &
      Match two images of the same animal \\[0.6ex]
    \texttt{sem\_map($l:X\!\to\! Y$)} &
      $\{\,M(l)(t) \mid t \in T\,\}$ &
      Extract brand names from product descriptions \\[0.6ex]
    \texttt{sem\_rank($l:T[X]\!\to\! Seq[X],\,k$)} &
      $\langle t_1,\dots,t_k\rangle$ ranked by $M(l)$ &
      Rank reviews by positivity \\[0.6ex]
    \texttt{sem\_classify($l:X\!\to\! Y$)} &
      $\{\, (t, M(l)(t)) \mid t \in T\,\}$ &
      Classify animals by species based on images \\
    \bottomrule
  \end{tabular}
\end{table*}


\subsection{Metrics}\label{subsec:metrics}


{ 
We evaluate SQPEs according to processing cost, result quality, and scalability. Processing cost includes execution time, the monetary cost of invoking LLMs, and memory usage. In our experience, costs according to at least one of these three metrics are the primary factors preventing us from using SQPEs on larger datasets. Result quality must also be evaluated, since LLM outputs may contain errors. Jointly measuring processing cost and result quality is similar to evaluation practices in existing work on approximate query processing (AQP)~\cite{ma2021learned, 10.14778/3648160.3648181, 10.1145/3725335}. This similarity arises because semantic query processing can be viewed as a form of AQP, since LLM outputs are inherently uncertain. As we evaluate analytical data processing engines, we focus on scaling with respect to the size of the processed dataset (rather than the number of users or concurrent queries), similar to prior benchmarks for analytical workloads (e.g., TPC-H).}

To assess result quality, we compare results generated by the SQPEs to the ground truth. We generate ground truth by exploiting the manual labels that come with our data sets (e.g., created for tasks such as classification). {We generate ground truth query results by substituting LLM calls with manual labels, provided in the data sets that our scenarios are based upon. For instance, to generate ground truth for a query counting positive movie reviews, we execute an SQL query that replaces the semantic predicate in the SemBench query, using the LLM to filter out negative reviews, with a simple SQL predicate, returing true if the review is assigned to a positive label in the underlying Kaggle data set (containing movie reviews with manual sentiment labels).}

The metric used to measure the result accuracy depends on the query type. For aggregation queries, calculating SQL aggregates such as counts or sums, we measure the relative distance between the ground truth aggregate value and the value returned by an SQPE. For retrieval queries, we calculate the F1 score, considering recall as the ratio of ground truth result values contained in the generated result, and precision as the ratio of rows in the generated result that appear in the ground truth result as well. For ranking queries, we use Spearman's rank correlation coefficient to measure the strength of correlation between the ranking produced by an SQPE and the ranking derived from ground-truth scores. For queries involving grouping operations—such as classification, simple mappings, or certain types of joins—we use the Adjusted Rand Index (ARI), a standard clustering similarity metric, to assess accuracy.


%% file: sections/04_scenarios.tex
\section{Benchmark Scenarios}
\label{sec:scenarios}

In the following subsections, we describe each of our benchmark scenarios in more detail, focusing on both, workload and data. In Section~\ref{sub:ScenarioComparison}, we compare all scenarios according to multiple criteria.

\subsection{Movies}

This scenario analyzes and compares movie reviews using sentiment analysis. Since text is the most broadly supported modality, the scenario is constructed to evaluate a wide range of systems and semantic operators using only tabular and textual content. We use movie review data from Kaggle~\cite{andrezaza2023moviereview} and construct two tables:

\begin{lstlisting}[
language=SQL,
showspaces=false,
basicstyle=\small\ttfamily,
numbers=left,
numberstyle=\tiny,
commentstyle=\color{gray}
]
Movies(id text, title text)
Reviews(id text, reviewId text, reviewText text)
\end{lstlisting}

The {\tt Movies} table contains metadata for each movie, and the {\tt Reviews} table stores critic reviews. In this scenario, our queries are designed to evaluate a large number of systems with various semantic operators, including semantic filter (e.g., selecting five positive reviews for a given movie), semantic join (e.g., checking whether two reviews express the same or opposite sentiment), semantic classification (e.g., classifying movie reviews into different sentiments), and semantic ranking (e.g., ranking reviews based on the degree of positivity).

\begin{example}
The following query counts the number of positive reviews for the movie \texttt{taken\_3}.
\begin{lstlisting}[
language=SQL,
showspaces=false,
numbers=left,
numberstyle=\tiny,
basicstyle=\small\ttfamily,
commentstyle=\color{gray}
]
SELECT COUNT(*) AS positive_review_cnt
FROM Reviews R
WHERE R.id = 'taken_3'
AND AI.IF(R.reviewText, 'this review expresses a positive sentiment');
\end{lstlisting}
\end{example}

\subsection{Wildlife}

This scenario aims at identifying the presence and co-occurrence of animal species, based on camera trap pictures (we use a Kaggle data set containing camera trap pictures with species labels~\cite{rushibalajiputthewad2024animalsoundclassify}) and audio recordings (similarly, we use a Kaggle data set~\cite{hypnotu2023dsailporiniwildlifeimagedataset, mugambi2023dsailwildlifeimagepaper} containing animal sound recordings with associated species labels). We generate a database with two tables:

\begin{lstlisting}[
  language=SQL,
  showspaces=false,
  basicstyle=\small\ttfamily,
  numbers=left,
  numberstyle=\tiny,
  commentstyle=\color{gray}
]
ImageTable(image img, city text, stationID text)
AudioTable(audio audio, city text, StationID text)
\end{lstlisting}

Both tables combine the image or audio recording with a randomly generated station ID and city (five possible cities and four possible stations). In this scenario, our queries focus in particular on semantic filters on audio and image data.

\begin{example}
The following query determines cities where elephants are present, indicated either by a corresponding picture or audio recording.
\begin{lstlisting}[
  language=SQL,
  showspaces=false,
  basicstyle=\small\ttfamily,
  numbers=left,
  numberstyle=\tiny,
  commentstyle=\color{gray}
]
SELECT DISTINCT city 
FROM (SELECT city FROM ImageTable I WHERE 
 AI.IF(I.image, 'the picture shows an elephant')) 
UNION (SELECT city FROM AudioTable  A WHERE 
 AI.IF(A.audio, 'this sounds like an elephant'));
\end{lstlisting}
\end{example}

\subsection{E-Commerce}

The E-Commerce scenario is inspired by an online fashion retail store, with the underlying dataset being available on Kaggle~\cite{param_aggarwal_2019}.
The dataset covers textual data, e.g., product description, structured data, as well as image data for each item, and overall comprises 44446 items, which also corresponds to the maximum scale factor. Products include various clothes as well as shoes, accessories, personal care, and home equipment.
The ground truth for each of the queries is established using the columns with structured data, e.g., the primary color of the item, the type of item etc.


\begin{lstlisting}[
  language=SQL,
  showspaces=false,
  basicstyle=\small\ttfamily,
  numbers=left,
  numberstyle=\tiny,
  commentstyle=\color{gray}
  ]
styles_details(id text, productDescription text, productImage img ...)
\end{lstlisting}

The queries of the E-Commerce scenario can be split into two groups, each with a separate goal: Single operator queries, which allow testing individual operators in isolation, as well as a set of complex queries with multiple operators. 


Queries 1-9 assess individual semantic operator performance in an isolated way across systems on textual and image modality.
Specifically, this includes the operators \texttt{SEM\_FILTER}, \texttt{SEM\_MAP}, and \texttt{SEM\_CLASSIFY} on textual and image data respectively, as well as \texttt{SEM\_JOIN} on text-to-text, text-to-image, and image-to-image joins.
This allows establishing a fundamental understanding of operator performances between different systems as well as assessing the impact of new ways of implementing operators in a standardized and easy-to-understand setting.
The prompts in these queries are deliberately kept simple with respect to modern LLM-standards to primarily focus on how efficiently a system can interact with a model and not on model-performance itself.


Queries 10-14 are more complex, requiring multiple different semantic operators on multiple modalities, oftentimes two different modalities as input to a single operator, and include many opportunities for complex cross-operator optimizations and other query optimization techniques, including prompt simplification and join ordering.

Currently, not all systems support all of these complex queries---in parts due to not supporting all required operators, in other parts due to excessive query runtimes even with a small scale factor (500).

\begin{example}
The following query finds product IDs of Reebok backpacks using semantic filtering on product descriptions.
\begin{lstlisting}[
  language=SQL,
  showspaces=false,
  basicstyle=\small\ttfamily,
  numbers=left,
  numberstyle=\tiny,
  commentstyle=\color{gray}
  ]
SELECT id
FROM styles_details
WHERE AI.IF('This is a backpack from Reebok',
               productDescription);
\end{lstlisting}
\end{example}

\subsection{Cars}\label{subsec:cars}

\textcolor{black}{
This scenario simulates a multi-modal car damage system to assist manufacturers and insurance companies with investigating damage presence and co-occurrence through diverse data modalities. We construct four interconnected tables based on four publicly available, anonymized datasets from Kaggle~\cite{hendrichs_cullen_2023, Krause_2013_ICCV_Workshops, kaggle_car_dignostics, cavazos_nhtsa_complaints}}:
\begin{lstlisting}[
  language=SQL,
  showspaces=false,
  basicstyle=\small\ttfamily,
  numbers=left,
  numberstyle=\tiny,
  commentstyle=\color{gray}
]
Cars(CarId int, Mileage float, FuelType text, TransmissionType text, VIN int, RegistrationDate text, Country text, NumberPlate text, PreviousOwners int)
ComplaintsText(CarId int, ComplaintId int, 
  Complaint text)
CarImage(CarId int, ImageId int,  Image image)
DiagnostAudio(CarId int, AudioId int, Audio audio)
\end{lstlisting}

\textcolor{black}{
The cars scenario spans all four modalities available in our benchmark (tables, text, images, and audio). It combines synthetic tabular vehicle records with complementary real-world multi-modal data: first-person problem descriptions (text), diagnostic recordings (audio), and images of intact and damaged cars.
A given car may exhibit zero, one, or multiple issues. Multi-modal inputs can, therefore, reflect either damage or a normal state. When possible, the same damage is represented across compatible modalities, e.g., a car with brake damage may include both abnormal brake sounds in the diagnostic audio and visible brake-related damage in images. Cars may also present distinct damages across modalities, e.g., a dent visible in images and an airbag deployment mentioned in text.
}

\textcolor{black}{
To construct the tabular component, we synthesize non-visual vehicle attributes and randomly assign damages under a set of co-occurrence constraints. For each car, we sample a manufacturer year (2000–2025), mileage (0–200k), fuel type (e.g., gasoline, hybrid, electric), transmission (automatic, manual, CVT), and a country from a broad list. We then generate identifiers and registration metadata, including a 17-character alphanumeric VIN and a registration date. License plates are produced using country-specific format templates, including localized character sets where applicable. Finally, we assign the number of previous owners using a weighted probabilistic model that increases the expected owner count for older and higher-mileage vehicles.
For the non-tabular modalities, we utilize existing real-world datasets. During evaluation, we treat the original labels from these source datasets as ground truth that is not exposed to the evaluated system.
}

\begin{example}
\textcolor{black}{The following query determines all damaged cars.}
\begin{lstlisting}[
  language=SQL,
  showspaces=false,
  numbers=left,
  numberstyle=\tiny,
  basicstyle=\small\ttfamily,
  commentstyle=\color{gray}
]
(SELECT CarId FROM ComplaintsText WHERE 
 AI.IF('The car is damaged according to the symptoms.', Complaint)) 
 UNION DISTINCT
(SELECT CarId FROM DiagnostAudio WHERE 
 AI.IF('The car is damaged according to the audio.',  Audio)) 
 UNION DISTINCT
(SELECT CarId FROM CarImage WHERE 
 AI.IF('The car is damaged according to the image.', Image));
\end{lstlisting}
\label{example:cars_query}
\end{example}

\subsection{MMQA}
The MMQA scenario simulates a multi-modal question answering system and is built on top of the MultiModalQA dataset~\cite{MmqaTalmor2021}. It covers three modalities including tables, texts and images. For this scenario, we construct five tables from Wikipedia tables and texts within the MultiModalQA dataset, along with an additional metadata table for the images from the same dataset. We also curate 11 questions that involve various AI operators such as semantic filters, joins, extraction and summarization.

\begin{lstlisting}[
  language=SQL,
  showspaces=false,
  basicstyle=\small\ttfamily,
  numbers=left,
  numberstyle=\tiny,
  commentstyle=\color{gray}
]
ap_warrior(id int, finish varchar, race varchar, distance varchar, track varchar, ...)
ben_piazza(year int, title varchar, role varchar, notes varchar)
ben_piazza_text_data(row_id int,title varchar,url varchar, id varchar, text varchar)
lizzy_caplan_text_data(row_id int, title varchar, url varchar, text varchar)
tampa_international_airport(row_id int , airlines varchar, destinations varchar, airport varchar)
images(row_id int, image_filename varchar, image_filepath varchar)
\end{lstlisting}

\begin{example}
The following query joins airlines with images containing their logos.
\begin{lstlisting}[
  language=SQL,
  showspaces=false,
  numbers=left,
  numberstyle=\tiny,
  basicstyle=\small\ttfamily,
  commentstyle=\color{gray}
]
SELECT t.Airlines, i.uri
FROM mmqa.tampa_international_airport t, mmqa.images i
WHERE AI.IF("You will be provided with an airline name and an image. Determine if the image shows the logo of the airline. Airline: ", t.Airlines, ", Image: ", i.uri);
\end{lstlisting}
\end{example}

\subsection{Comparison of Scenarios}
\label{sub:ScenarioComparison}

\begin{table*}[t]
  \centering
  \caption{Comparison of Different Scenarios.}
  \label{tab:scenario-stats}
  \setlength\tabcolsep{4.4pt}
  \begin{tabular}{l r cccc ccccc ccc}
    \toprule
    \multirow{2}{*}{\textbf{Scenario}} &
    \multirow{2}{*}{\textbf{\#Queries}} &
    \multicolumn{4}{c}{\textbf{Modalities}} &
    \multicolumn{5}{c}{\textbf{Number of Semantic Operators}} &
    \multicolumn{3}{c}{\textbf{Modality Sizes (\#rows)}} \\
    \cmidrule(lr){3-6} \cmidrule(lr){7-11} \cmidrule(lr){12-14}
     & & \textbf{Table} & \textbf{Text} & \textbf{Image} & \textbf{Audio} 
       & \textbf{Filter} & \textbf{Join} & \textbf{Map} & \textbf{Rank} & \textbf{Classify} 
       & \textbf{Text} & \textbf{Image} & \textbf{Audio} \\
    \midrule
    Movie & 10 & \checkmark & \checkmark & -- & -- 
          & 4 & 3 & -- & 2 & 1 
          & 1,375,738 & -- & -- \\
    Wildlife & 10 & \checkmark & -- & \checkmark & \checkmark 
          & 17 & -- & -- & -- & -- 
          & -- & \phantom{0}8,718 & 650 \\
    E-Commerce & 14 & \checkmark & \checkmark & \checkmark & -- 
          & 12 & 9 & 3 & 1  & 2
          & \phantom{000}44,446 & 44,446 & -- \\
    MMQA & 11 & \checkmark & \checkmark & \checkmark & -- 
          & 5 & 3 & 4 & --  & --
          & \phantom{0000}5,000 & \phantom{0}1,000 & -- \\
    \textcolor{black}{Cars} & \textcolor{black}{10} & \checkmark & \checkmark & \checkmark & \checkmark 
          & \textcolor{black}{12} & -- & -- & --  & \textcolor{black}{1}
          & \phantom{0}\textcolor{black}{157,376} & \textcolor{black}{30,131} & \textcolor{black}{1,387} \\
    \midrule
    \textbf{Total} & \textbf{55} & \checkmark & \checkmark & \checkmark & \checkmark 
          & \textbf{50} & \textbf{15} & \textbf{7} & \textbf{3} & \textbf{4}
          & \textbf{1,582,560} & \textbf{84,295} & \textbf{2037} \\

    \bottomrule
  \end{tabular}%
\end{table*}

Table~\ref{tab:scenario-stats} compares all of the scenarios introduced before. The scenarios are complementary in terms of the data types to which semantic operators are applied. \textcolor{black}{Since LLMs perform differently across domains, we select five representative scenarios, based on common academic datasets and inspired by workloads observed in industry, to cover a broad range of data domains. 
As SQPEs expand their support for semantic operators and data modalities, and as LLMs face new challenges in emerging data domains, we plan to release additional scenarios in future SemBench instantiations.}

The \textcolor{black}{Cars} scenario is most diverse in terms of data modalities, applying semantic operators to text, images, and audio files. On the other hand, the Movies scenario is more restricted and applies semantic operators only to analyze text data. The Movies scenario enables developers to use our benchmark in parts for evaluating SQPEs that only support a limited set of modalities (typically, text).

Different scenarios focus on different semantic operators. The semantic filter operator, supported by all of the SQPEs we tested, is used in all scenarios. The E-Commerce scenario is the most join-heavy, using 9 semantic join operators in its 10 benchmark queries. Three scenarios use semantic classification, whereas two scenarios use semantic ranking and the semantic map operator, respectively.

Additionally, Table~\ref{tab:scenario-stats} shows the number of labeled items for each data modality (text, images, and audio files) that appear in each benchmark data set. Note that data processing via semantic operators is much more expensive than data processing with relational operators. With current systems, processing even a small part of our benchmark data sets is prohibitively expensive for many of our queries. Therefore, we only use a small part of our data (a few hundred to a few thousand rows, depending on the scenario) for the experiments. While the number of rows may seem small, compared to the row counts typically used to evaluate systems on benchmarks such as TPC-H, it is largely sufficient to evaluate SQPEs.

%% file: sections/06_experimental_results.tex
\section{Experimental Results}
\label{sec:experiments}

In Section~\ref{sub:ExperimentalSetup}, we describe our general experimental setup. In Section~\ref{sub:SystemSettings}, we discuss how each of the evaluated systems was tuned. In Section~\ref{subsec:results}, we report and analyze our experimental results. \textcolor{black}{In Section~\ref{subsubsec:scalability}, we present scalability experiments. In Section~\ref{subsubsec:trade-offs}, we compare and discuss different systems.}

\subsection{Experimental Setup}
\label{sub:ExperimentalSetup}

\paragraphnospace{Hardware Settings.} \textcolor{black}{The experiments in Section~\ref{subsec:results}} are conducted on an EC2 instance of type \texttt{g4dn.2xlarge}, configured with 32 GB of RAM, 8 vCPUs, an NVIDIA T4 GPU with 16 GB of memory, and 250 GB of disk storage. The GPU is used solely for embedding computation, and we do not deploy LLMs locally. \textcolor{black}{The scalability experiments are conducted on an Ubuntu server with two Intel Xeon Gold 5218 CPUs and 384~GB of RAM. We use a different server because current academic systems require significantly more memory than the EC2 instance can provide when scaling up (more than 200~GB for some systems). More details are discussed in Section~\ref{subsubsec:memory}.}


\paragraphnospace{Model Settings.}
For all experiments presented in the paper, we use \texttt{gemini-2.5-flash}, which belongs to the latest Gemini-series, as the backing LLM across all systems.
The choice of the model has the biggest influence on system performance---hence it is crucial to evaluate all systems with the same model.
We chose \texttt{gemini-2.5-flash} because it is available for use in all evaluated systems and supports all required modalities in the benchmark.
Our experiments also showed that the model strikes a good balance between cost and output quality in the context of large-scale multimodal data processing.
Additionally, \textcolor{black}{we} disable the reasoning capability of the LLM in the main experiments due to long processing times and high monetary cost and, as we see in the Appendix, not necessarily better results. We further set the model temperature to $0$ for applicable systems for reproducibility.
Results for additional models can be found on our online website at {\url{https://sembench.org}}.




\paragraphnospace{Degree of Parallelism.}
We set the degree of parallelism for LLM invocation to 20 for all applicable systems. 
BigQuery does not give users the option to control this setting.

\paragraphnospace{Scenario Settings.}
\textcolor{black}{
To control the database size, we define a \textit{Scale} for each scenario as the number of rows in the primary entity table, with other table sizes defined as functions of it. This is similar to the scaling factors in TPC-H~\cite{TPC2013} and other benchmarks~\cite{tpcds} for studying system behavior at different scales.
Table~\ref{tab:sf-experiments} lists the available Scale ranges and the Scales used in our experiments. Base scales are chosen based on scalability tests to balance system stress with feasible monetary cost for future benchmark users. We also evaluate larger Scales with scalability experiments in Section~\ref{subsubsec:scalability}, where a single query can exceed one hundred dollars, 200 GB main memory, and current systems do not complete within a reasonable time. In total, the cost of LLM calls when running the scalability experiments is \$9,935.80, taking approximately 18 days to execute. As shown in Table~\ref{tab:scenario-stats}, the full corpus contains over 1.5 million text rows, 84 thousand image rows, and two thousand audio rows, making the largest Scales impractical today. As systems improve, higher Scales can be used in future evaluations.
}

\begin{table}[t]
  \centering
  \caption{Scales for Different Scenarios.}
  \label{tab:sf-experiments}
  \setlength\tabcolsep{6pt}
  \begin{tabular}{lccccc}
    \toprule
    \textbf{Scenario} & \textbf{Available Range} & \multicolumn{2}{c}{\textbf{Scale Used}} \\
    \cmidrule(lr){3-4}
     &  & \textbf{Base} & \textcolor{black}{\textbf{Scalability}} \\
    \midrule
    Movie       & 1--1,375,738 & 2,000  & \textcolor{black}{1,000--16,000} \\
    Wildlife    & 1--8,718     & 200    & \textcolor{black}{100--1,600} \\
    E-Commerce  & 50--44,446   & 500    & \textcolor{black}{250--4,000} \\
    \textcolor{black}{Cars}
                & \textcolor{black}{1--157,376}
                                & \textcolor{black}{19,672}
                                         & \textcolor{black}{9,836--157,376} \\
    MMQA        & 25--1,000    & 200    & \textcolor{black}{100--800} \\
    \bottomrule
  \end{tabular}
\end{table}

\paragraphnospace{Evaluation Metrics.}
We report monetary cost, result quality, and query latency in our evaluation. For quality metrics, we use the metrics described in Section~\ref{subsec:metrics} in our experiments. All metrics are normalized to the range $[0,1]$, where higher values indicate better performance. This normalization ensures consistency in both visualization and comparison. Specifically, the F1 score naturally falls within $[0,1]$. For relative error, we apply the transformation $1 / (1 + \text{relative\_error})$ to map it into $[0,1]$. Spearman’s rank correlation~\cite{spearmanCoeff} and the Adjusted Rand Index (ARI) originally range from $[-1,1]$, where $-1$ indicates opposite ranking or clustering similarity, $1$ indicates identical outcomes, and $0$ corresponds to random behavior. In our experiments, all evaluated systems produce rank correlation and ARI values greater than $0$. Therefore, we do not perform additional normalization for these two metrics.



\paragraphnospace{Evaluation Settings.} Each experiment is repeated five times. We report the average performance of monetary cost, quality, and latency per query, as well as the overall averages across all queries. In the tables, we also report the standard deviation across all queries. 

\subsection{System Settings}
\label{sub:SystemSettings}


\textcolor{black}{All systems build the final prompt by combining prompt templates with a natural language operator configuration provided as part of the semantic query. The prompt templates define operator semantics and vary across systems to reflect different design choices, such as reducing token usage or improving output quality. SemBench does not change the default templates used by different systems. Hence, even when using the same operator configurations in the queries sent to different systems, the prompts sent to the LLM may differ due to different prompt templates. For all systems, we use default settings or author-recommended configurations for all tuning parameters. 
}


\paragraphnospace{LOTUS.} We use LOTUS version \texttt{1.1.3} for our experiments. LOTUS provides options for optimized semantic filter, join, and rank. In evaluation, we fix the model to \texttt{gemini-2.5-flash}. For semantic join, we use the optimized operator from LOTUS, where embedding similarity scores are used as approximations to LLM calls. Following the recommendations of the authors, we use \texttt{e5-base-v2} for text embeddings and \texttt{clip-ViT-B-32} for image embeddings. 


\paragraphnospace{Palimpzest.}
We use Palimpzest version \texttt{0.8.2} with the Abacus optimizer~\cite{russo2025abacus} turned off. We apply the MaxQuality objective, which configures Palimpzest to use its default ``best'' operator implementation(s) based on prior belief. 
Model selection is disabled since a fixed model is used in the evaluation.

\paragraphnospace{ThalamusDB.} We use version \texttt{0.1.15} and the default settings for all parameters. In particular, we
set the batch size for the join operator to ten (i.e., ten rows from both input tables are integrated into the same prompt). Specifically for the \textcolor{black}{Cars} scenario, we slightly rewrote queries to eliminate SQL WITH clauses, materializing the query results described in the WITH clause as temporary tables.

\paragraphnospace{BigQuery.} In the evalutaion, we use a default reservation of 2000 slots and fix the model to \texttt{gemini-2.5-flash}. Scalability parameters such as LLM parallelism or batch size are managed by the engine and are not controllable by users.

\subsection{Detailed Results per Scenario}\label{subsec:results}




\textcolor{black}{Tables~\ref{tab:experimental_results_all} reports the results for the five scenarios.} 
The colors indicate the relative performance of the different systems for each query. 
For cost and latency, we normalize the values using 
\((v - \min) / (\max - \min)\), and classify the systems into three categories: 
the top 33\% are labeled \textit{Better}, the middle 33\% are labeled \textit{Middle}, and the bottom 33\% are labeled \textit{Worse}. 
For quality, we use absolute thresholds: values in \([0.8, 1]\) are labeled \textit{Better}, 
values in \([0.6, 0.8)\) are labeled \textit{Middle}, and values in \([0, 0.6)\) are labeled \textit{Worse}. 
The left-most column shows the query number together with the semantic operators used in the query.
It also reports per-query averages for each system along with the \textcolor{black}{absolute} standard deviation \textcolor{black}{of the average (i.e., denoting by $\sigma_i$ the standard deviation of query $i$ out of $n$ queries, we report $\sqrt{\sum_i\sigma_i^2}/n$)}. The standard deviation is small for cost and quality compared to run time. Variance in cost and quality mainly comes from randomness in LLM outputs, which systems often reduce by using a low temperature when possible. In contrast, run time variability is largely due to fluctuations in LLM call latency.




\input{tables/academics/combined_table_1}
\input{tables/academics/combined_table_2}




For the Movies scenario, most systems support all queries. ThalamusDB does not support the last two queries, requiring semantic rank operators, which are not supported by ThalamusDB. Despite a small scale of 500, we find some queries (e.g., Q7)  require several minutes of processing time, along with non-negligible monetary cost, for most of the compared systems. This shows that even relatively small amounts of data compared to database standards pose significant challenges in semantic query processing.



\textcolor{black}{Comparing only systems supporting all queries}, BigQuery, the only commercial system in our evaluation, achieves optimal performance according to all metrics, followed closely by LOTUS (in terms of cost) and Palimpzest (in terms of quality). \textcolor{black}{While ThalamusDB only supports a subset of queries in this scenario, it achieves optimal costs for each supported query (eight out of ten).}

Interestingly, there are several queries for which the relative cost difference between different systems is significant. 
For Queries Q1, Q5, and Q6, processing costs vary by more than a factor of $100\times$ across different systems. All three queries use a LIMIT clause, but the systems handle it differently. In LOTUS, the LIMIT clause is applied only as a post-processing step after semantic operators are evaluated on all data. In contrast, SPQEs can terminate semantic operator evaluation early once enough rows are produced to satisfy the LIMIT clause, which leads to significantly better performance for these queries. Query Q2 uses a LIMIT clause as well, but restricts the scope to reviews of one specific movie (using SQL standard equality predicates, which are evaluated efficiently). In this case, applying semantic operators to all relevant data is less costly than for the other three LIMIT queries (which consider all reviews).

Among all queries without LIMIT clause, Q7 is the one with the most significant cost differences. ThalamusDB achieves minimal average processing costs, followed by LOTUS. This is explained by the fact that these two systems use implementations of the semantic join operators that are aimed at reducing execution costs. LOTUS implements semantic joins by matching items first based on their embedding vectors, evaluating the join condition via LLM calls only on those row pairs whose embedding vectors are within a certain distance range. ThalamusDB uses batching to include multiple items from both tables into the prompt, instructing the LLM to find all matches regarding the join condition. On the other hand, those cost optimization strategies come at the expense of quality. BigQuery and Palimpzest achieve the highest average quality.

Table~\ref{tab:experimental_results_all} (b) shows results for the E-Commerce scenario.
This scenario features the most diverse selection of semantic operators, making it challenging to implement it for all systems even though all generally support the required data modalities.
LOTUS and BigQuery are the only systems that allow a complete implementation of this scenario, with Palimpzest being only missing out on one query because it does not support \texttt{SEM\_RANK} and also doesn't allow emulating \texttt{SEM\_RANK} with a \texttt{SEM\_MAP} together with a \texttt{ORDER\_BY} as the system does not support classical sorting.
ThalamusDB only supports 5 out of the 14 queries due to only supporing \texttt{SEM\_FILTER} and \texttt{SEM\_JOIN} together with the limitation that only a single input column for semantic operators is supported.
While Q10 and Q11 are implementable in BigQuery, they do not complete within a reasonable execution time.
Notably, LOTUS consistently achieves the lowest cost across all queries, being outperformed only by ThalamusDB on Q1 and Q7, while still achieving the best average accuracy in this scenario. \textcolor{black}{Note that the cost of LOTUS demonstrates the highest variance across all systems in this scenario. The reason lies in the implementation of the semantic join in LOTUS. LOTUS analyzes a data sample to determine thresholds on the distance between embedding vectors, within which candidate join pairs are evaluated via the LLM. Comparing different runs, that sample, and therefore the thresholds, differ, meaning that the number of join candidates processed by the LLM varies significantly.}






Table~\ref{tab:experimental_results_all} (c) shows results for the Wildlife scenario. This scenario uses semantic operators on audio data and images. LOTUS does not currently support analysis of audio data. Hence, it only supports four out of ten queries in this scenario. The other systems support all ten queries. \textcolor{black}{Among the systems supporting all queries}, ThalamusDB achieves the lowest average costs. \textcolor{black}{Note that the standard deviation for cost is close to zero in this scenario. This is due to the fact that queries use exclusively semantic filter operators. For this operator and all compared systems, the cost is very stable across different runs. Unlike other systems, BigQuery shows a relatively high quality variance across different runs. This behavior arises because the model temperature is set to zero for the other systems, whereas model-related parameters in BigQuery are controlled internally and are not exposed to users. As a result, BigQuery exhibits higher output variability. BigQuery achieves the highest average quality among all systems supporting all queries (with a gap of over three standard deviations).}

Cost differences are generally smaller in the Wildlife scenario, with relative differences of up to a factor of ten, compared to more than two orders of magnitude in the Movie scenario. The Wildlife uses only the semantic filter operator and no semantic joins, which typically cause larger cost differences across systems. Analyzing the compared open-source systems, we find significant differences in the prompt templates used to implement the semantic filter. All systems use simple and compact prompt templates, especially for output tokens to reduce costs. Palimpzest uses relatively long prompt templates, containing examples for few-shot learning and detailed instructions. This increases quality while increasing costs associated with reading input tokens. ThalamusDB uses concise prompt templates that reduce costs at the expense of quality.

Table~\ref{tab:experimental_results_all} (d) presents the experimental results for the MMQA scenario. Only Palimpzest fully supports MMQA queries. LOTUS can technically support all semantic operators, but it may produce internal processing errors across runs. These errors occur when LLM responses do not follow the output schema specified in the prompt and the subsequent code does not handle unexpected outputs. This issue explains the ``n/a'' entry for Q4, despite LOTUS supporting the semantic map operator. ThalamusDB only supports a small number of queries, in particular due to the lack of a semantic map operator and a semantic join operator across modalities. While BigQuery can perform one-to-one semantic mapping, they lack the many-to-one semantic map operator, which is essential for summarizing values across multiple rows within a single column. 

\textcolor{black}{Among the systems that successfully handled at least nine out of ten queries}, BigQuery proved to be the most cost-effective solution. In contrast, Palimpzest achieved the highest average quality, with LOTUS following closely behind. Across all queries, the most challenging and expensive operations---in terms of both monetary cost and latency---were semantic joins between tables and images. This difficulty is illustrated by the results for Q7, which required 40,000 LLM calls for join predicate evaluation. This single query took Palimpzest over 35 minutes and LOTUS nearly 40 minutes to complete. The monetary cost for this query alone was more than 1,000 times that of a filtering query. The inherent complexity of these joins is also reflected in their consistently lower quality scores compared to other operations like semantic filtering and mapping.

\begin{figure*}[thbp]
    \centering
    \includegraphics[width=0.82\linewidth]{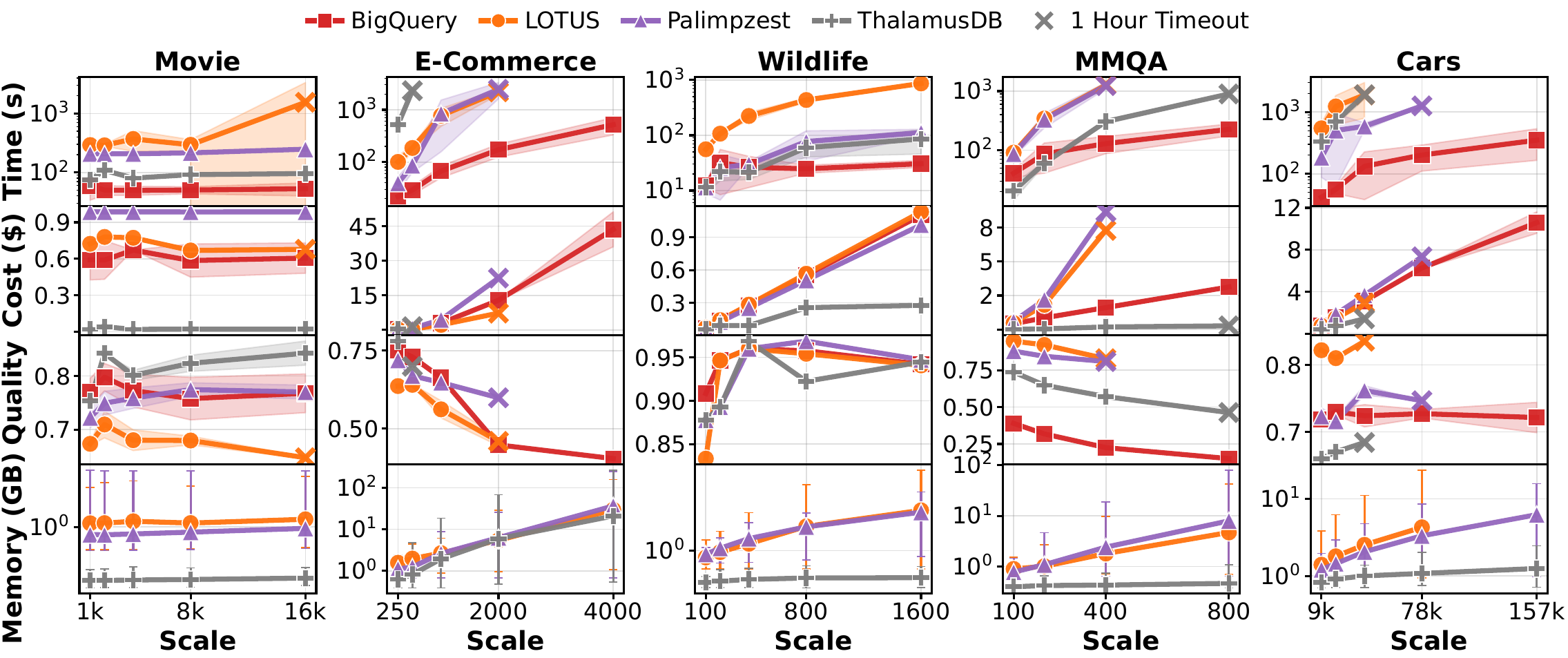}

    \caption{\textcolor{black}{Scalability across scenarios. We report average execution time, monetary cost, result quality, and memory usage over a fixed set of queries that are supported by all systems. Error bars for execution time, cost, and result quality show standard deviations over five runs. For memory usage, error bars show the minimum and maximum values. Cross markers indicate a per-query timeout of 1 hour. After a timeout, the system is treated as failed and is not evaluated at larger scales.}}

    \label{fig:scalabillity_memory}
    \Description{}
\end{figure*}

\textcolor{black}{
Table~\ref{tab:experimental_results_all} (e) presents the results for the Cars scenario.
First, current systems demonstrate poor performance in car diagnostics audio processing. Q2, Q5, and Q6, which rely on semantic filtering of audio data, are either unsupported by LOTUS or perform poorly across systems.
Second, existing systems offer limited operator support. 
ThalamusDB lacks semantic classification required for Q10 and inconsistently supports common table expressions, necessitating the manual materialization of temporary tables for Q6. 
Palimpzest does not support relational joins and requires us to hardcode the join order and execute the joins with Pandas (Q6 and Q7).
Third, Q5, Q6, and Q7, which process image data, exhibit similar quality across all evaluated systems, while the execution costs differ significantly, with Palimpzest being the most expensive. We attribute this to the higher computational cost of image semantic operators relative to text-only queries.
Fourth, in the case of semantic operators for textual damage detection, queries targeting cars with more specialized damages (e.g., Q4 - engine problems) yield higher quality results compared to those involving more generic conditions (e.g., Q1 - crash/accident/collision). We hypothesize that LLMs perform better when the problem space is narrower and more well-defined.
The results indicate that none of the evaluated systems are yet capable of adequately addressing all queries in the specialized cars scenario.
Unlike in traditional databases, the quality of results produced by semantic operators is highly dependent on the domain and modality of the data being processed, and systems need to be aware of this fact and actively manage it.
}

\subsection{\textcolor{black}{Scalability Experiments}}\label{subsubsec:scalability}

\textcolor{black}{We evaluate system scalability by creating multiple databases for each scenario that vary the data size around a scenario-specific Base Scale: \{0.5$\times$ Base, Base, 2$\times$ Base, 4$\times$ Base, 8$\times$ Base\}, and the scales used are summarized in Table~\ref{tab:sf-experiments}. For each system and scenario, evaluation starts from the smallest scale with a per-query timeout of one hour. If any query exceeds this limit at a given scale, all remaining queries at that scale are still executed, but the system is not evaluated at larger scales. Systems are not forcibly terminated upon timeout, and executions that complete beyond one hour are still recorded. At each scale, results are averaged over the set of common queries supported by all systems. As shown in Figure~\ref{fig:scalabillity_memory}, we report average execution time, monetary cost, and result quality as functions of the scale factor (the bottom row reports memory consumption in a separate scalability experiment and is discussed later). Error bars indicate standard deviations across five runs, and the one-hour timeout is marked using cross markers. Due to space limitations, we provide the full per-query results for all scales on our online website at \url{https://sembench.org}.}

\textcolor{black}{For the Movie scenario, nearly all systems complete each query within 1 hour for all scales. Unlike other scenarios that require more resources to process images and audio data, this scenario uses only text data. LOTUS reaches the timeout of one hour at scale 16K due to Q6, which uses a semantic join with a LIMIT clause. Unlike other systems, LOTUS does not terminate early after collecting the number of rows required by the LIMIT clause, but continues processing data. This increases cost and time overheads for Q6.}

\textcolor{black}{
For the E-Commerce and MMQA scenarios, all systems except BigQuery fail to scale to large scales. BigQuery scales to these scales because it has an internal mechanism to adjust resources, such as the number of parallel LLM invocations, based on input size. This mechanism is not available in the other systems.
The two scenarios include expensive multi-modal semantic join operators, and the number of LLM calls for these operators grows quadratically with table size. For example, for Q7 in the MMQA scenario, a scale of 400 results in 160,000 LLM calls, which leads to monetary costs of about \$50 for LOTUS and \$60 for Palimpzest. Similarly, Q8 in the E-Commerce scenario uses a semantic join over text and images. Only BigQuery can scale to the largest scale of 4000 for this query, with a cost of about \$140.
These results point to an important optimization problem: how to scale multi-modal semantic joins over large tables. In addition, as data sizes increase, result quality consistently decreases across all systems. Therefore, scaling processing while maintaining high result quality remains an open problem.
}

\textcolor{black}{
For the Wildlife scenario, all systems scale to the largest scale because this scenario does not include semantic join operators, but only semantic filters over images and audio data. First, the monetary cost of BigQuery increases linearly with the scale, while the execution time does not. As discussed earlier, this behavior is due to BigQuery adjusting resources for each query based on input size, a feature not available in other systems. Second, ThalamusDB achieves much lower monetary cost because it avoids unnecessary LLM evaluations during query execution. Taking Q5 from the Wildlife scenario as an example, Q5 asks for cities that have either images or audio recordings of elephants. Once ThalamusDB finds evidence that a city satisfies this condition, it stops evaluating additional rows for that city, since further evaluations cannot change the final result. In contrast, other systems continue to evaluate all relevant rows. Third, we find that result quality increases slightly for some queries as we scale up. Taking Q7 as an example, this query asks for cities where zebras and impalas co-occur. As the scale factor increases, more data is sampled, and the number of rows supporting each city also increases. With more supporting rows, systems have more chances to successfully evaluate at least one correct row for each city, which increases the likelihood of returning the correct result.
}

\textcolor{black}{
Finally, Cars is similar to Wildlife in terms of operators and modalities, as it does not include semantic joins but uses semantic filter and classification operators over text, image, and audio data. For Cars, we are able to increase the scale to cover all data. In this scenario, BigQuery, Palimpzest, ThalamusDB, and LOTUS rank from best to worst in execution time.
Execution time and monetary cost increase nearly linearly with scale, while BigQuery achieves lower execution time due to its resource adjustment mechanism. The result quality remains stable across different scales.
}

\input{tables/aggregate_table_common}


\label{subsubsec:memory} \textcolor{black}{
We conduct one run of a separate experiment without a timeout to measure memory usage at scale, and Figure~\ref{fig:scalabillity_memory} (fourth row) reports the average, minimum, and maximum peak memory usage per query for each scenario. 
BigQuery memory usage is not reported because this information is not available to users. The average memory usage increases linearly with the scale factor. 
In contrast, the maximum memory usage varies widely across scenarios. It is below or around 10~GB in the Movie, Wildlife, and Cars scenarios, but reaches hundreds of gigabytes in the E-Commerce and MMQA scenarios. 
For example, 
Q8 in the E-Commerce scenario requires about 250~GB, 150~GB, and 100~GB of memory for ThalamusDB, LOTUS, and Palimpzest, respectively.
These systems create image pairs to evaluate semantic joins and send them to remote LLMs. The memory used to store these image pairs is not released, so memory usage accumulates as more pairs are processed. This leads to significant resource waste and can cause server crashes. In addition, the memory usage of LOTUS on the Cars dataset at a scale of 157k is not reported. LOTUS does not handle LLM invocation failures robustly and may stall for over 48 hours at large scale due to network issues, unstable API, or rate limiting. This result indicates that systems should explicitly handle LLM invocation failures.
}




\subsection{\textcolor{black}{System Comparison}}\label{subsubsec:trade-offs}
\noindent \textcolor{black}{
Table~\ref{tab:aggregate_operator_results} shows aggregated results (for the base scale) for query groups that contain specific semantic operators. All comparative claims are based on a gap between average values of at least three standard deviations. 
For Filter, Map, Score, and Classify, where current implementations mainly differ in prompt design, LOTUS and Palimpzest attain the best quality by using detailed prompt templates, but this choice leads to higher costs due to increased input tokens. ThalamusDB uses concise prompt templates that result in the lowest costs (0.19 for Filter) but lower quality (0.698 for Filter). For semantic joins, cost reduction strategies introduce additional trade-offs. Palimpzest uses a naive semantic join implementation that generates item pairs for LLM evaluation in a nested loop manner. LOTUS applies embedding-based pre-filtering to reduce the number of LLM calls. This reduces the monetary cost to \$2.74, compared to \$3.29 for Palimpzest. However, pre-filtering can miss valid matches and leads to lower quality, with a score of 0.614 compared to 0.703. ThalamusDB uses batching to process multiple row pairs within a single prompt and asks LLMs to decide which row pairs satisfy the join condition. This reduces the cost to \$0.14, but it yields lower quality of 0.487, compared to 0.703 for Palimpzest. Overall, the systems achieve different quality–cost trade-offs through different design choices. ThalamusDB achieves the lowest average cost, at only \$0.17 on average, but with lower average quality. BigQuery achieves the best latency, with an average latency of 48.1s, and provides a good balance between quality and cost. LOTUS and Palimpzest achieve higher quality, but at moderately higher cost.
}

\textcolor{black}{Finally, we report several lessons learned from our experiments. First, a system should exploit LIMIT clauses efficiently, terminating early once the required number of rows is generated. LOTUS does not use this technique and must process the entire data set before applying post-processing, increasing cost. Second, a system should decide during query execution whether additional rows need to be evaluated. ThalamusDB supports early termination when further processing cannot change the final result. Third, a system should be able to adjust the degree of parallelism at runtime based on the sizes of the input relations, as done in BigQuery. Fourth, LLM invocation failures are common due to network issues, unstable APIs, or rate limits; therefore, systems should handle these failures explicitly by using retries with exponential backoff. Finally, SQPEs should limit memory consumption when processing multimodal data. For instance, LOTUS, Palimpzest, and ThalamusDB all encounter memory pressure when processing images with semantic joins, limiting their scalability.}


%% file: tables/academics/combined_table_1.tex
\begin{table*}[htbp]
  \centering
  \caption{SemBench Results. Operators: F - semantic filter, J - semantic join, R - semantic rank, C - semantic classify, L - LIMIT clause. Colors: \protect\fancycellGreen{Better than Average}, \protect\fancycellYellow{Average}, \protect\fancycellGray{Worse than Average}, and \protect\fancyCellRed{Not supported} for not supported.}
  \label{tab:experimental_results_all}

  \begingroup
  \normalsize
  \setlength{\tabcolsep}{3.7pt}
  \setlength{\aboverulesep}{0.0ex}
  \setlength{\belowrulesep}{0.0ex}
  \renewcommand{\arraystretch}{\arraystretchresulttable}

  \begin{tabular*}{\textwidth}{@{\extracolsep{\fill}} l
      c@{\hspace{0.25em}}c@{\hspace{0.25em}}c@{\hspace{1em}}
      c@{\hspace{0.25em}}c@{\hspace{0.25em}}c@{\hspace{1em}}
      c@{\hspace{0.25em}}c@{\hspace{0.25em}}c@{\hspace{1em}}
      c@{\hspace{0.25em}}c@{\hspace{0.25em}}c}

    \SemBenchHeader
    
    \ScenarioRow{(a) Movie Scenario}
        \textbf{Q1:} F L & \fancycellGray{\$0.09} & \fancycellGreen{1.00} & \fancycellGray{33.1\,\text{s}} & \fancycellGreen{\$1${\cdot\scriptstyle 10^{-3}}$} & \fancycellGreen{1.00} & \fancycellGreen{3.8\,\text{s}} & \fancycellGreen{\$4${\cdot\scriptstyle 10^{-4}}$} & \fancycellGreen{0.95} & \fancycellGreen{4.2\,\text{s}} & \fancycellYellow{\$0.05} & \fancycellGreen{1.00} & \fancycellGray{26.3\,\text{s}} \\
    \textbf{Q2:} F L & \fancycellGray{\$0.01} & \fancycellGreen{1.00} & \fancycellGreen{2.1\,\text{s}} & \fancycellGray{\$0.01} & \fancycellGreen{1.00} & \fancycellGray{29.7\,\text{s}} & \fancycellGreen{\$2${\cdot\scriptstyle 10^{-3}}$} & \fancycellGreen{0.92} & \fancycellGreen{1.9\,\text{s}} & \fancycellGreen{\$3${\cdot\scriptstyle 10^{-3}}$} & \fancycellGreen{1.00} & \fancycellGreen{9.5\,\text{s}} \\
    \textbf{Q3:} F & \fancycellYellow{\$5${\cdot\scriptstyle 10^{-3}}$} & \fancycellYellow{0.64} & \fancycellGreen{2.1\,\text{s}} & \fancycellGray{\$0.01} & \fancycellYellow{0.64} & \fancycellGreen{4.6\,\text{s}} & \fancycellGreen{\$2${\cdot\scriptstyle 10^{-3}}$} & \fancycellYellow{0.74} & \fancycellGreen{3.1\,\text{s}} & \fancycellGreen{\$3${\cdot\scriptstyle 10^{-3}}$} & \fancycellYellow{0.64} & \fancycellGray{11.0\,\text{s}} \\
    \textbf{Q4:} F & \fancycellGreen{\$5${\cdot\scriptstyle 10^{-3}}$} & \fancycellYellow{0.64} & \fancycellGreen{2.8\,\text{s}} & \fancycellGray{\$0.02} & \fancycellYellow{0.74} & \fancycellGreen{4.4\,\text{s}} & \fancycellGreen{\$2${\cdot\scriptstyle 10^{-3}}$} & \fancycellYellow{0.74} & \fancycellGreen{3.8\,\text{s}} & \fancycellGreen{\$3${\cdot\scriptstyle 10^{-3}}$} & \fancycellYellow{0.64} & \fancycellGray{11.4\,\text{s}} \\
    \textbf{Q5:} J L & \fancycellGray{\$2.38} & \fancycellGray{0.59} & \fancycellGray{536.5\,\text{s}} & \fancycellGreen{\$0.01} & \fancycellGray{0.39} & \fancycellGreen{1.9\,\text{s}} & \fancycellGreen{\$1${\cdot\scriptstyle 10^{-3}}$} & \fancycellGreen{1.00} & \fancycellGreen{2.3\,\text{s}} & \fancycellYellow{\$1.01} & \fancycellGreen{0.89} & \fancycellGreen{54.5\,\text{s}} \\
    \textbf{Q6:} J L & \fancycellGray{\$1.81} & \fancycellYellow{0.67} & \fancycellGray{432.4\,\text{s}} & \fancycellGreen{\$0.01} & \fancycellGreen{0.83} & \fancycellGreen{2.3\,\text{s}} & \fancycellGreen{\$9${\cdot\scriptstyle 10^{-4}}$} & \fancycellGreen{0.84} & \fancycellGreen{1.7\,\text{s}} & \fancycellYellow{\$1.00} & \fancycellYellow{0.69} & \fancycellGreen{54.5\,\text{s}} \\
    \textbf{Q7:} J & \fancycellGreen{\$1.81} & \fancycellGray{0.21} & \fancycellGreen{431.8\,\text{s}} & \fancycellGray{\$7.72} & \fancycellYellow{0.68} & \fancycellGray{1056.1\,\text{s}} & \fancycellGreen{\$0.15} & \fancycellGray{0.57} & \fancycellYellow{636.9\,\text{s}} & \fancycellYellow{\$3.31} & \fancycellYellow{0.70} & \fancycellGreen{198.3\,\text{s}} \\
    \textbf{Q8:} C & \fancycellGreen{\$4${\cdot\scriptstyle 10^{-3}}$} & \fancycellGreen{0.93} & \fancycellGreen{2.3\,\text{s}} & \fancycellGray{\$0.02} & \fancycellGreen{0.86} & \fancycellGreen{4.3\,\text{s}} & \fancycellGreen{\$5${\cdot\scriptstyle 10^{-3}}$} & \fancycellGreen{0.83} & \fancycellYellow{6.8\,\text{s}} & \fancycellGreen{\$3${\cdot\scriptstyle 10^{-3}}$} & \fancycellYellow{0.76} & \fancycellGray{10.9\,\text{s}} \\
    \textbf{Q9:} R & \fancycellGreen{\$0.02} & \fancycellYellow{0.75} & \fancycellGreen{4.9\,\text{s}} & \fancycellGray{\$0.05} & \fancycellYellow{0.78} & \fancycellGreen{5.7\,\text{s}} & \fancyCellRed{n/a} & \fancyCellRed{n/a} & \fancyCellRed{n/a} & \fancycellGreen{\$0.02} & \fancycellYellow{0.78} & \fancycellGray{13.3\,\text{s}} \\
    \textbf{Q10:} R & \fancycellGreen{\$0.13} & \fancycellGray{0.40} & \fancycellGreen{30.9\,\text{s}} & \fancycellGray{\$0.38} & \fancycellGray{0.42} & \fancycellGray{39.2\,\text{s}} & \fancyCellRed{n/a} & \fancyCellRed{n/a} & \fancyCellRed{n/a} & \fancycellGreen{\$0.13} & \fancycellGray{0.44} & \fancycellGreen{32.1\,\text{s}} \\
    \addlinespace[0.3ex]
    \midrule
    \addlinespace[0.3ex]
    \textbf{Avg} & \fancycellNormal{\$0.62} & \fancycellNormal{0.68} & \fancycellNormal{147.9\,\text{s}} & \fancycellNormal{\$0.82} & \fancycellNormal{0.73} & \fancycellNormal{115.2\,\text{s}} & \fancycellNormal{\$0.02} & \fancycellNormal{0.82} & \fancycellNormal{82.6\,\text{s}} & \fancycellNormal{\$0.55} & \fancycellNormal{0.75} & \fancycellNormal{42.2\,\text{s}} \\
        \textcolor{black}{\textbf{Std Dev}} &
\textcolor{black}{$\pm\text{\$}0.02$} &
\textcolor{black}{$\pm 0.01$} &
\textcolor{black}{$\pm 5.0\text{s}$} &
\textcolor{black}{$\pm\text{\$}3{\cdot\scriptstyle 10^{-5}}$} &
\textcolor{black}{$\pm 0.01$} &
\textcolor{black}{$\pm 12.2\text{s}$} &
\textcolor{black}{$\pm\text{\$}6{\cdot\scriptstyle 10^{-5}}$} &
\textcolor{black}{$\pm 0.00$} &
\textcolor{black}{$\pm 1.3\text{s}$} &
\textcolor{black}{$\pm\text{\$}0.01$} &
\textcolor{black}{$\pm 0.01$} &
\textcolor{black}{$\pm 6.5\text{s}$} \\
    \bottomrule
    
    \midrule
    \ScenarioRow{(b) E-Commerce Scenario}
     \textbf{Q1:} F & \fancycellYellow{\$0.06} & \fancycellGreen{1.00} & \fancycellGreen{12.2\,\text{s}} & \fancycellGray{\$0.08} & \fancycellGreen{1.00} & \fancycellGreen{12.2\,\text{s}} & \fancycellGreen{\$0.03} & \fancycellGreen{1.00} & \fancycellGray{34.8\,\text{s}} & \fancycellGreen{\$0.04} & \fancycellGray{0.59} & \fancycellYellow{21.2\,\text{s}} \\
    \textbf{Q2:} F & \fancycellGreen{\$0.18} & \fancycellGreen{0.87} & \fancycellGray{166.1\,\text{s}} & \fancycellGreen{\$0.38} & \fancycellGreen{0.83} & \fancycellGreen{57.5\,\text{s}} & \fancycellGreen{\$0.31} & \fancycellYellow{0.67} & \fancycellYellow{100.8\,\text{s}} & \fancycellGray{\$3.96} & \fancycellGray{0.21} & \fancycellGreen{55.7\,\text{s}} \\
    \textbf{Q3:} M & \fancycellGreen{\$0.07} & \fancycellGreen{0.97} & \fancycellGreen{17.1\,\text{s}} & \fancycellGray{\$0.12} & \fancycellGreen{0.98} & \fancycellGreen{16.5\,\text{s}} & \fancyCellRed{n/a} & \fancyCellRed{n/a} & \fancyCellRed{n/a} & \fancycellGray{\$0.12} & \fancycellGreen{0.97} & \fancycellGray{21.2\,\text{s}} \\
    \textbf{Q4:} M & \fancycellGreen{\$0.14} & \fancycellGray{0.45} & \fancycellYellow{156.7\,\text{s}} & \fancycellYellow{\$0.24} & \fancycellGray{0.53} & \fancycellGray{335.0\,\text{s}} & \fancyCellRed{n/a} & \fancyCellRed{n/a} & \fancyCellRed{n/a} & \fancycellGray{\$0.37} & \fancycellYellow{0.69} & \fancycellGreen{31.0\,\text{s}} \\
    \textbf{Q5:} C & \fancycellGreen{\$0.04} & \fancycellGreen{0.99} & \fancycellGreen{7.4\,\text{s}} & \fancycellGreen{\$0.07} & \fancycellGreen{0.98} & \fancycellGreen{6.6\,\text{s}} & \fancyCellRed{n/a} & \fancyCellRed{n/a} & \fancyCellRed{n/a} & \fancycellGray{\$0.17} & \fancycellGreen{0.98} & \fancycellGray{25.7\,\text{s}} \\
    \textbf{Q6:} C & \fancycellGreen{\$0.12} & \fancycellGreen{0.89} & \fancycellGray{114.8\,\text{s}} & \fancycellGray{\$0.43} & \fancycellGreen{0.89} & \fancycellGray{143.7\,\text{s}} & \fancyCellRed{n/a} & \fancyCellRed{n/a} & \fancyCellRed{n/a} & \fancycellGray{\$0.35} & \fancycellGreen{0.88} & \fancycellGreen{34.9\,\text{s}} \\
    \textbf{Q7:} J & \fancycellGray{\$1.33} & \fancycellYellow{0.75} & \fancycellYellow{199.4\,\text{s}} & \fancycellGray{\$1.79} & \fancycellGreen{0.92} & \fancycellGray{287.6\,\text{s}} & \fancycellGreen{\$0.08} & \fancycellGray{0.51} & \fancycellGreen{97.7\,\text{s}} & \fancycellYellow{\$0.86} & \fancycellGreen{0.83} & \fancycellGreen{45.4\,\text{s}} \\
    \textbf{Q8:} J & \fancycellGreen{\$4${\cdot\scriptstyle 10^{-3}}$} & \fancycellGreen{1.00} & \fancycellGreen{4.1\,\text{s}} & \fancycellGreen{\$0.01} & \fancycellGreen{1.00} & \fancycellGreen{3.9\,\text{s}} & \fancycellGreen{\$0.30} & \fancycellGray{0.00} & \fancycellGray{713.0\,\text{s}} & \fancycellGray{\$18.23} & \fancycellGray{0.29} & \fancycellGreen{126.2\,\text{s}} \\
    \textbf{Q9:} J & \fancycellGreen{\$0.06} & \fancycellGray{0.55} & \fancycellGreen{243.4\,\text{s}} & \fancycellGreen{\$0.10} & \fancycellGray{0.49} & \fancycellGreen{44.9\,\text{s}} & \fancycellGray{\$0.31} & \fancycellGray{0.00} & \fancycellGray{872.2\,\text{s}} & \fancycellYellow{\$0.21} & \fancycellGray{0.58} & \fancycellGreen{48.6\,\text{s}} \\
    \textbf{Q10:} F J & \fancycellGreen{\$0.21} & \fancycellGray{0.00} & \fancycellGreen{519.5\,\text{s}} & \fancycellGray{\$1.02} & \fancycellGray{0.06} & \fancycellGray{1192.5\,\text{s}} & \fancyCellRed{n/a} & \fancyCellRed{n/a} & \fancyCellRed{n/a} & \fancyCellRed{n/a} & \fancyCellRed{n/a} & \fancyCellRed{n/a} \\
    \textbf{Q11:} F J & \fancycellGreen{\$0.27} & \fancycellYellow{0.78} & \fancycellGray{158.7\,\text{s}} & \fancycellGray{\$0.61} & \fancycellYellow{0.73} & \fancycellGreen{132.4\,\text{s}} & \fancyCellRed{n/a} & \fancyCellRed{n/a} & \fancyCellRed{n/a} & \fancyCellRed{n/a} & \fancyCellRed{n/a} & \fancyCellRed{n/a} \\
    \textbf{Q12:} F M & \fancycellGreen{\$0.10} & \fancycellYellow{0.60} & \fancycellGray{36.4\,\text{s}} & \fancycellGray{\$0.14} & \fancycellGray{0.00} & \fancycellGreen{31.9\,\text{s}} & \fancyCellRed{n/a} & \fancyCellRed{n/a} & \fancyCellRed{n/a} & \fancycellGreen{\$0.10} & \fancycellGreen{0.97} & \fancycellGreen{31.1\,\text{s}} \\
    \textbf{Q13:} F & \fancycellGreen{\$0.24} & \fancycellYellow{0.74} & \fancycellGray{274.8\,\text{s}} & \fancycellGray{\$0.44} & \fancycellYellow{0.74} & \fancycellGray{238.3\,\text{s}} & \fancyCellRed{n/a} & \fancyCellRed{n/a} & \fancyCellRed{n/a} & \fancycellGray{\$0.38} & \fancycellYellow{0.70} & \fancycellGreen{22.4\,\text{s}} \\
    \textbf{Q14:} F J R & \fancycellGreen{\$0.23} & \fancycellGreen{0.87} & \fancycellGray{178.2\,\text{s}} & \fancyCellRed{n/a} & \fancyCellRed{n/a} & \fancyCellRed{n/a} & \fancyCellRed{n/a} & \fancyCellRed{n/a} & \fancyCellRed{n/a} & \fancycellGray{\$4.26} & \fancycellGray{0.37} & \fancycellGreen{73.6\,\text{s}} \\
    \addlinespace[0.3ex]
    \midrule
    \addlinespace[0.3ex]
    \textbf{Avg} & \fancycellNormal{\$0.22} & \fancycellNormal{0.75} & \fancycellNormal{149.2\,\text{s}} & \fancycellNormal{\$0.42} & \fancycellNormal{0.70} & \fancycellNormal{192.5\,\text{s}} & \fancycellNormal{\$0.21} & \fancycellNormal{0.44} & \fancycellNormal{363.7\,\text{s}} & \fancycellNormal{\$2.42} & \fancycellNormal{0.67} & \fancycellNormal{44.7\,\text{s}} \\
     \textcolor{black}{\textbf{Std Dev}} &
\textcolor{black}{$\pm\text{\$}0.03$} &
\textcolor{black}{$\pm 0.01$} &
\textcolor{black}{$\pm 35.4\text{s}$} &
\textcolor{black}{$\pm\text{\$}4{\cdot\scriptstyle 10^{-4}}$} &
\textcolor{black}{$\pm 0.01$} &
\textcolor{black}{$\pm 101.0\text{s}$} &
\textcolor{black}{$\pm\text{\$}1{\cdot\scriptstyle 10^{-5}}$} &
\textcolor{black}{$\pm 0.00$} &
\textcolor{black}{$\pm 71.0\text{s}$} &
\textcolor{black}{$\pm\text{\$}0.01$} &
\textcolor{black}{$\pm 0.03$} &
\textcolor{black}{$\pm 3.4\text{s}$} \\
    \bottomrule

    \midrule
    \ScenarioRow{(c) Wildlife Scenario}
    \textbf{Q1:} F & \fancycellGreen{\$0.11} & \fancycellYellow{0.79} & \fancycellGray{92.4\,\text{s}} & \fancycellGray{\$0.13} & \fancycellYellow{0.79} & \fancycellGreen{32.8\,\text{s}} & \fancycellGreen{\$0.11} & \fancycellYellow{0.79} & \fancycellGreen{19.6\,\text{s}} & \fancycellGreen{\$0.11} & \fancycellYellow{0.79} & \fancycellGreen{32.0\,\text{s}} \\
    \textbf{Q2:} F & \fancyCellRed{n/a} & \fancyCellRed{n/a} & \fancyCellRed{n/a} & \fancycellYellow{\$0.01} & \fancycellGray{0.17} & \fancycellGreen{2.8\,\text{s}} & \fancycellYellow{\$0.01} & \fancycellGray{0.14} & \fancycellGreen{4.5\,\text{s}} & \fancycellYellow{\$0.01} & \fancycellGray{0.19} & \fancycellGray{9.4\,\text{s}} \\
    \textbf{Q3:} F L & \fancycellGray{\$0.11} & \fancycellGreen{1.00} & \fancycellGray{99.4\,\text{s}} & \fancycellGray{\$0.13} & \fancycellGray{0.00} & \fancycellGreen{22.7\,\text{s}} & \fancycellGreen{\$0.03} & \fancycellGray{0.00} & \fancycellGreen{4.3\,\text{s}} & \fancycellGray{\$0.11} & \fancycellGray{0.00} & \fancycellGreen{25.7\,\text{s}} \\
    \textbf{Q4:} F L & \fancyCellRed{n/a} & \fancyCellRed{n/a} & \fancyCellRed{n/a} & \fancycellGray{\$0.01} & \fancycellGray{0.00} & \fancycellGreen{2.7\,\text{s}} & \fancycellGreen{\$1${\cdot\scriptstyle 10^{-3}}$} & \fancycellGray{0.00} & \fancycellGreen{1.3\,\text{s}} & \fancycellGray{\$0.01} & \fancycellGreen{1.00} & \fancycellGray{9.4\,\text{s}} \\
    \textbf{Q5:} F & \fancyCellRed{n/a} & \fancyCellRed{n/a} & \fancyCellRed{n/a} & \fancycellGray{\$0.13} & \fancycellYellow{0.75} & \fancycellYellow{13.5\,\text{s}} & \fancycellGreen{\$0.01} & \fancycellYellow{0.75} & \fancycellGreen{2.3\,\text{s}} & \fancycellGray{\$0.12} & \fancycellYellow{0.75} & \fancycellGray{19.2\,\text{s}} \\
    \textbf{Q6:} F & \fancyCellRed{n/a} & \fancyCellRed{n/a} & \fancyCellRed{n/a} & \fancycellGray{\$0.13} & \fancycellGray{0.00} & \fancycellYellow{19.3\,\text{s}} & \fancycellGreen{\$0.06} & \fancycellGray{0.50} & \fancycellGreen{13.2\,\text{s}} & \fancycellGray{\$0.12} & \fancycellGray{0.20} & \fancycellGray{24.3\,\text{s}} \\
    \textbf{Q7:} F & \fancycellGray{\$0.23} & \fancycellGreen{1.00} & \fancycellGray{188.3\,\text{s}} & \fancycellGreen{\$0.13} & \fancycellGreen{1.00} & \fancycellGreen{43.9\,\text{s}} & \fancycellGray{\$0.20} & \fancycellGreen{1.00} & \fancycellGreen{28.8\,\text{s}} & \fancycellGray{\$0.22} & \fancycellGreen{1.00} & \fancycellGreen{24.6\,\text{s}} \\
    \textbf{Q8:} F & \fancyCellRed{n/a} & \fancyCellRed{n/a} & \fancyCellRed{n/a} & \fancycellGreen{\$0.13} & \fancycellYellow{0.75} & \fancycellGray{34.9\,\text{s}} & \fancycellGreen{\$0.12} & \fancycellYellow{0.75} & \fancycellGreen{23.8\,\text{s}} & \fancycellGray{\$0.23} & \fancycellYellow{0.75} & \fancycellGray{35.5\,\text{s}} \\
    \textbf{Q9:} F & \fancyCellRed{n/a} & \fancyCellRed{n/a} & \fancyCellRed{n/a} & \fancycellGray{\$0.13} & \fancycellGray{0.57} & \fancycellGreen{19.1\,\text{s}} & \fancycellGreen{\$0.08} & \fancycellYellow{0.67} & \fancycellGreen{17.2\,\text{s}} & \fancycellGray{\$0.12} & \fancycellGray{0.59} & \fancycellGray{37.6\,\text{s}} \\
    \textbf{Q10:} F L & \fancycellGray{\$0.11} & \fancycellGreen{1.00} & \fancycellGray{87.8\,\text{s}} & \fancycellGray{\$0.13} & \fancycellGray{0.00} & \fancycellGreen{19.6\,\text{s}} & \fancycellGreen{\$0.03} & \fancycellGray{0.00} & \fancycellGreen{4.4\,\text{s}} & \fancycellGray{\$0.11} & \fancycellGray{0.00} & \fancycellYellow{40.3\,\text{s}} \\
    \addlinespace[0.3ex]
    \midrule
    \addlinespace[0.3ex]
    \textbf{Avg} & \fancycellNormal{\$0.14} & \fancycellNormal{0.95} & \fancycellNormal{117.0\,\text{s}} & \fancycellNormal{\$0.10} & \fancycellNormal{0.40} & \fancycellNormal{21.1\,\text{s}} & \fancycellNormal{\$0.07} & \fancycellNormal{0.46} & \fancycellNormal{11.9\,\text{s}} & \fancycellNormal{\$0.11} & \fancycellNormal{0.53} & \fancycellNormal{25.8\,\text{s}} \\
    \textcolor{black}{\textbf{Std Dev}} &
\textcolor{black}{$\pm\text{\$}7{\cdot\scriptstyle 10^{-18}}$} &
\textcolor{black}{$\pm 0.00$} &
\textcolor{black}{$\pm 6.2\text{s}$} &
\textcolor{black}{$\pm\text{\$}5{\cdot\scriptstyle 10^{-18}}$} &
\textcolor{black}{$\pm 0.00$} &
\textcolor{black}{$\pm 5.2\text{s}$} &
\textcolor{black}{$\pm\text{\$}1{\cdot\scriptstyle 10^{-18}}$} &
\textcolor{black}{$\pm 0.00$} &
\textcolor{black}{$\pm 0.51\text{s}$} &
\textcolor{black}{$\pm\text{\$}2{\cdot\scriptstyle 10^{-18}}$} &
\textcolor{black}{$\pm 0.02$} &
\textcolor{black}{$\pm 7.6\text{s}$} \\
    \bottomrule

    \midrule
    \ScenarioRow{(d) MMQA Scenario}
     \textbf{Q1:} M & \fancycellGray{\$0.02} & \fancycellGreen{1.00} & \fancycellGreen{7.4\,\text{s}} & \fancycellGreen{\$3${\cdot\scriptstyle 10^{-3}}$} & \fancycellGreen{1.00} & \fancycellGreen{4.5\,\text{s}} & \fancyCellRed{n/a} & \fancyCellRed{n/a} & \fancyCellRed{n/a} & \fancycellYellow{\$0.01} & \fancycellGreen{1.00} & \fancycellGray{14.1\,\text{s}} \\
    \textbf{Q2a:} J & \fancycellGray{\$0.89} & \fancycellGreen{0.83} & \fancycellGray{169.6\,\text{s}} & \fancycellGray{\$1.04} & \fancycellGreen{1.00} & \fancycellGray{135.9\,\text{s}} & \fancyCellRed{n/a} & \fancyCellRed{n/a} & \fancyCellRed{n/a} & \fancycellGreen{\$0.08} & \fancycellGray{0.00} & \fancycellGreen{53.8\,\text{s}} \\
    \textbf{Q2b:} J & \fancycellGray{\$0.89} & \fancycellGreen{0.83} & \fancycellGray{166.8\,\text{s}} & \fancycellGray{\$1.04} & \fancycellGreen{1.00} & \fancycellGray{133.8\,\text{s}} & \fancyCellRed{n/a} & \fancyCellRed{n/a} & \fancyCellRed{n/a} & \fancycellGreen{\$0.12} & \fancycellGray{0.00} & \fancycellGreen{38.4\,\text{s}} \\
    \textbf{Q3a:} F & \fancycellGreen{\$0.01} & \fancycellGreen{0.83} & \fancycellGreen{4.2\,\text{s}} & \fancycellGray{\$0.02} & \fancycellGreen{0.80} & \fancycellGreen{4.5\,\text{s}} & \fancycellGreen{\$0.01} & \fancycellYellow{0.75} & \fancycellYellow{11.0\,\text{s}} & \fancycellGreen{\$0.01} & \fancycellYellow{0.72} & \fancycellGray{19.6\,\text{s}} \\
    \textbf{Q3f:} F & \fancycellGreen{\$0.01} & \fancycellGreen{1.00} & \fancycellGreen{4.2\,\text{s}} & \fancycellGray{\$0.02} & \fancycellGreen{1.00} & \fancycellGreen{8.0\,\text{s}} & \fancycellGreen{\$0.01} & \fancycellGreen{1.00} & \fancycellGreen{7.0\,\text{s}} & \fancycellGreen{\$0.01} & \fancycellYellow{0.67} & \fancycellGray{22.3\,\text{s}} \\
    \textbf{Q4:} M & \fancyCellRed{n/a} & \fancyCellRed{n/a} & \fancyCellRed{n/a} & \fancycellGray{\$5${\cdot\scriptstyle 10^{-3}}$} & \fancycellGray{0.54} & \fancycellGreen{1.2\,\text{s}} & \fancyCellRed{n/a} & \fancyCellRed{n/a} & \fancyCellRed{n/a} & \fancycellGreen{\$2${\cdot\scriptstyle 10^{-3}}$} & \fancycellYellow{0.60} & \fancycellGray{9.7\,\text{s}} \\
    \textbf{Q5:} M & \fancycellGreen{\$1${\cdot\scriptstyle 10^{-3}}$} & \fancycellGreen{1.00} & \fancycellGreen{0.48\,\text{s}} & \fancycellGray{\$1${\cdot\scriptstyle 10^{-3}}$} & \fancycellGreen{1.00} & \fancycellGray{0.50\,\text{s}} & \fancyCellRed{n/a} & \fancyCellRed{n/a} & \fancyCellRed{n/a} & \fancyCellRed{n/a} & \fancyCellRed{n/a} & \fancyCellRed{n/a} \\
    \textbf{Q6a:} F & \fancycellYellow{\$0.01} & \fancycellGreen{1.00} & \fancycellGreen{4.4\,\text{s}} & \fancycellGray{\$0.02} & \fancycellGreen{1.00} & \fancycellGreen{4.0\,\text{s}} & \fancycellGreen{\$2${\cdot\scriptstyle 10^{-3}}$} & \fancycellGray{0.33} & \fancycellGreen{5.6\,\text{s}} & \fancycellGreen{\$4${\cdot\scriptstyle 10^{-3}}$} & \fancycellGray{0.03} & \fancycellGray{18.9\,\text{s}} \\
    \textbf{Q6b:} F & \fancycellYellow{\$0.01} & \fancycellGreen{1.00} & \fancycellGreen{3.8\,\text{s}} & \fancycellGray{\$0.02} & \fancycellGreen{1.00} & \fancycellGreen{4.1\,\text{s}} & \fancycellGreen{\$2${\cdot\scriptstyle 10^{-3}}$} & \fancycellGreen{1.00} & \fancycellGreen{5.5\,\text{s}} & \fancycellGreen{\$4${\cdot\scriptstyle 10^{-3}}$} & \fancycellGray{0.04} & \fancycellGray{17.3\,\text{s}} \\
    \textbf{Q6c:} F & \fancycellYellow{\$0.01} & \fancycellGreen{1.00} & \fancycellGreen{4.6\,\text{s}} & \fancycellGray{\$0.02} & \fancycellGreen{1.00} & \fancycellGreen{4.7\,\text{s}} & \fancycellGreen{\$2${\cdot\scriptstyle 10^{-3}}$} & \fancycellGray{0.53} & \fancycellGray{13.4\,\text{s}} & \fancycellGreen{\$4${\cdot\scriptstyle 10^{-3}}$} & \fancycellGray{0.13} & \fancycellGray{17.4\,\text{s}} \\
    \textbf{Q7:} J & \fancycellGray{\$13.61} & \fancycellGray{0.32} & \fancycellGray{2311.4\,\text{s}} & \fancycellGray{\$15.65} & \fancycellGray{0.31} & \fancycellGray{2101.6\,\text{s}} & \fancyCellRed{n/a} & \fancyCellRed{n/a} & \fancyCellRed{n/a} & \fancycellGreen{\$1.18} & \fancycellGray{0.00} & \fancycellGreen{91.7\,\text{s}} \\
    \addlinespace[0.3ex]
    \midrule
    \addlinespace[0.3ex]
            \textbf{Avg} & \fancycellNormal{\$1.41} & \fancycellNormal{0.88} & \fancycellNormal{243.4\,\text{s}} & \fancycellNormal{\$1.62} & \fancycellNormal{0.88} & \fancycellNormal{218.4\,\text{s}} & \fancycellNormal{\$5${\cdot\scriptstyle 10^{-3}}$} & \fancycellNormal{0.72} & \fancycellNormal{8.5\,\text{s}} & \fancycellNormal{\$0.14} & \fancycellNormal{0.32} & \fancycellNormal{30.3\,\text{s}} \\
   \textcolor{black}{\textbf{Std Dev}} &
\textcolor{black}{$\pm\text{\$}8{\cdot\scriptstyle 10^{-7}}$} &
\textcolor{black}{$\pm 0.00$} &
\textcolor{black}{$\pm 2.3\text{s}$} &
\textcolor{black}{$\pm\text{\$}2{\cdot\scriptstyle 10^{-5}}$} &
\textcolor{black}{$\pm 0.00$} &
\textcolor{black}{$\pm 5.3\text{s}$} &
\textcolor{black}{$\pm\text{\$}0.00$} &
\textcolor{black}{$\pm 0.00$} &
\textcolor{black}{$\pm 2.7\text{s}$} &
\textcolor{black}{$\pm\text{\$}5{\cdot\scriptstyle 10^{-5}}$} &
\textcolor{black}{$\pm 0.00$} &
\textcolor{black}{$\pm 3.2\text{s}$} \\
\bottomrule

  \end{tabular*}
  \endgroup
\end{table*}

%% file: tables/academics/combined_table_2.tex
\begin{table*}[t]
  \ContinuedFloat
  \centering

  \begingroup
  \normalsize
  \setlength{\tabcolsep}{3.7pt}
  \setlength{\aboverulesep}{0.0ex}
  \setlength{\belowrulesep}{0.0ex}
  \renewcommand{\arraystretch}{\arraystretchresulttable}

  \begin{tabular*}{\textwidth}{@{\extracolsep{\fill}} l
      c@{\hspace{0.25em}}c@{\hspace{0.25em}}c@{\hspace{1em}}
      c@{\hspace{0.25em}}c@{\hspace{0.25em}}c@{\hspace{1em}}
      c@{\hspace{0.25em}}c@{\hspace{0.25em}}c@{\hspace{1em}}
      c@{\hspace{0.25em}}c@{\hspace{0.25em}}c}

    \ContinuedNote
    \midrule
    
    \ScenarioRow{\textcolor{black}{(e) Cars Scenario}}
    \textbf{Q1:} F & \fancycellYellow{\$1.74} & \fancycellGreen{0.90} & \fancycellYellow{550.0\,\text{s}} & \fancycellGray{\$2.44} & \fancycellYellow{0.69} & \fancycellYellow{465.6\,\text{s}} & \fancycellGreen{\$1.37} & \fancycellGreen{0.81} & \fancycellGray{829.7\,\text{s}} & \fancycellGreen{\$1.44} & \fancycellYellow{0.71} & \fancycellGreen{61.7\,\text{s}} \\
    \textbf{Q2:} F & \fancyCellRed{n/a} & \fancyCellRed{n/a} & \fancyCellRed{n/a} & \fancycellGreen{\$4${\cdot\scriptstyle 10^{-3}}$} & \fancycellGray{0.00} & \fancycellGreen{4.1\,\text{s}} & \fancycellGray{\$0.01} & \fancycellGray{0.09} & \fancycellGray{14.1\,\text{s}} & \fancycellGray{\$0.01} & \fancycellGray{0.08} & \fancycellGray{14.1\,\text{s}} \\
    \textbf{Q3:} F L & \fancycellYellow{\$0.60} & \fancycellGreen{0.90} & \fancycellGray{456.2\,\text{s}} & \fancycellGreen{\$0.01} & \fancycellGreen{0.92} & \fancycellGreen{6.1\,\text{s}} & \fancyCellRed{n/a} & \fancyCellRed{n/a} & \fancyCellRed{n/a} & \fancycellGray{\$1.66} & \fancycellGreen{1.00} & \fancycellGreen{36.4\,\text{s}} \\
    \textbf{Q4:} F & \fancycellYellow{\$1.71} & \fancycellGreen{0.99} & \fancycellGray{822.0\,\text{s}} & \fancycellGray{\$2.41} & \fancycellGreen{0.99} & \fancycellYellow{443.6\,\text{s}} & \fancycellGreen{\$1.34} & \fancycellGreen{1.00} & \fancycellGray{768.8\,\text{s}} & \fancycellGreen{\$1.41} & \fancycellGreen{0.99} & \fancycellGreen{68.7\,\text{s}} \\
    \textbf{Q5:} F & \fancyCellRed{n/a} & \fancyCellRed{n/a} & \fancyCellRed{n/a} & \fancycellGreen{\$0.01} & \fancycellGreen{1.00} & \fancycellGreen{6.3\,\text{s}} & \fancycellGray{\$1.61} & \fancycellGreen{1.00} & \fancycellGray{483.7\,\text{s}} & \fancycellGray{\$1.47} & \fancycellGreen{1.00} & \fancycellGreen{58.9\,\text{s}} \\
    \textbf{Q6:} F J & \fancyCellRed{n/a} & \fancyCellRed{n/a} & \fancyCellRed{n/a} & \fancycellGray{\$2.51} & \fancycellGreen{0.96} & \fancycellYellow{427.3\,\text{s}} & \fancycellGreen{\$1.96} & \fancycellGreen{0.97} & \fancycellGray{775.4\,\text{s}} & \fancycellGreen{\$2.00} & \fancycellGreen{0.96} & \fancycellGreen{44.3\,\text{s}} \\
    \textbf{Q7:} F & \fancyCellRed{n/a} & \fancyCellRed{n/a} & \fancyCellRed{n/a} & \fancycellGray{\$4.47} & \fancycellGray{0.56} & \fancycellYellow{882.7\,\text{s}} & \fancycellGreen{\$3.06} & \fancycellGray{0.58} & \fancycellGray{2146.2\,\text{s}} & \fancycellGreen{\$3.17} & \fancycellGray{0.45} & \fancycellGreen{86.0\,\text{s}} \\
    \textbf{Q8:} F L & \fancycellGray{\$1.78} & \fancycellGray{0.45} & \fancycellGray{1349.8\,\text{s}} & \fancycellGray{\$2.02} & \fancycellGray{0.29} & \fancycellGreen{268.7\,\text{s}} & \fancycellGreen{\$0.21} & \fancycellGray{0.20} & \fancycellGreen{68.0\,\text{s}} & \fancycellGray{\$1.69} & \fancycellGray{0.24} & \fancycellGreen{38.2\,\text{s}} \\
    \textbf{Q9:} F & \fancyCellRed{n/a} & \fancyCellRed{n/a} & \fancyCellRed{n/a} & \fancycellGray{\$2.05} & \fancycellGray{0.00} & \fancycellGray{308.1\,\text{s}} & \fancycellGray{\$1.61} & \fancycellGreen{1.00} & \fancycellGray{335.8\,\text{s}} & \fancycellGreen{\$0.03} & \fancycellGray{0.00} & \fancycellGreen{13.6\,\text{s}} \\
    \textbf{Q10:} C & \fancycellGreen{\$3.09} & \fancycellGray{0.41} & \fancycellGray{618.1\,\text{s}} & \fancycellGray{\$4.69} & \fancycellGray{0.51} & \fancycellGray{594.9\,\text{s}} & \fancyCellRed{n/a} & \fancyCellRed{n/a} & \fancyCellRed{n/a} & \fancycellGreen{\$2.70} & \fancycellGray{0.57} & \fancycellGreen{62.0\,\text{s}} \\
    \addlinespace[0.3ex]
    \midrule
    \addlinespace[0.3ex]
    \textbf{Avg} & \fancycellNormal{\$1.78} & \fancycellNormal{0.73} & \fancycellNormal{759.2\,\text{s}} & \fancycellNormal{\$2.06} & \fancycellNormal{0.59} & \fancycellNormal{340.7\,\text{s}} & \fancycellNormal{\$1.24} & \fancycellNormal{0.71} & \fancycellNormal{603.2\,\text{s}} & \fancycellNormal{\$1.56} & \fancycellNormal{0.60} & \fancycellNormal{48.4\,\text{s}} \\
\textcolor{black}{\textbf{Std Dev}} &
\textcolor{black}{$\pm\text{\$}4{\cdot\scriptstyle 10^{-17}}$} &
\textcolor{black}{$\pm 0.00$} &
\textcolor{black}{$\pm 92.5\text{s}$} &
\textcolor{black}{$\pm\text{\$}2{\cdot\scriptstyle 10^{-3}}$} &
\textcolor{black}{$\pm 0.01$} &
\textcolor{black}{$\pm 55.5\text{s}$} &
\textcolor{black}{$\pm\text{\$}5{\cdot\scriptstyle 10^{-17}}$} &
\textcolor{black}{$\pm 0.00$} &
\textcolor{black}{$\pm 14.3\text{s}$} &
\textcolor{black}{$\pm\text{\$}0.07$} &
\textcolor{black}{$\pm 0.00$} &
\textcolor{black}{$\pm 1.6\text{s}$} \\
    \bottomrule

    \bottomrule

    
  \end{tabular*}
  \endgroup
\end{table*}

%% file: tables/aggregate_table_common.tex
\begin{table*}[t]
  \centering
  \begin{threeparttable}
  \caption{\textcolor{black}{Aggregated Results by Operator Type. Colors indicate relative performance (\protect\fancycellGreen{Best than Average}, \protect\fancycellYellow{Average}, \protect\fancycellGray{Worse than Average}). For each operator, only queries supported by all available systems are included.}}
  \label{tab:aggregate_operator_results}

  \begingroup
  \small
  \setlength{\tabcolsep}{2pt}
  \setlength{\aboverulesep}{0.0ex}
  \setlength{\belowrulesep}{0.0ex}
  \renewcommand{\arraystretch}{\arraystretchresulttable}

  \begin{tabular*}{\textwidth}{@{\extracolsep{\fill}} l  c c c  c c c  c c c  c c c }
    \toprule
      & \multicolumn{3}{c}{\textbf{LOTUS}} & \multicolumn{3}{c}{\textbf{Palimpzest}} & \multicolumn{3}{c}{\textbf{ThalamusDB}} & \multicolumn{3}{c}{\textbf{BigQuery}} \\
      \cmidrule(lr){2-4} \cmidrule(lr){5-7} \cmidrule(lr){8-10} \cmidrule(lr){11-13}
      & \textbf{Quality} & \textbf{Latency} & \textbf{Cost} & \textbf{Quality} & \textbf{Latency} & \textbf{Cost} & \textbf{Quality} & \textbf{Latency} & \textbf{Cost} & \textbf{Quality} & \textbf{Latency} & \textbf{Cost} \\
    \midrule
    \addlinespace[0.3ex]
    \textbf{Filter} & \fancycellGreen{0.896} & \fancycellGray{194.4s} & \fancycellGreen{\$0.34} & \fancycellGreen{0.777} & \fancycellGreen{72.2s} & \fancycellGray{\$0.40} & \fancycellYellow{0.698} & \fancycellYellow{101.1s} & \fancycellGreen{\$0.19} & \fancycellGray{0.574} & \fancycellGreen{26.7s} & \fancycellYellow{\$0.36} \\
    \textbf{Join} & \fancycellGreen{0.614} & \fancycellGray{541.1s} & \fancycellYellow{\$2.74} & \fancycellGreen{0.703} & \fancycellYellow{454.3s} & \fancycellGray{\$3.29} & \fancycellGray{0.487} & \fancycellGreen{387.3s} & \fancycellGreen{\$0.14} & \fancycellYellow{0.601} & \fancycellGreen{69.4s} & \fancycellGreen{\$1.02} \\
    \textbf{Map} & \fancycellYellow{0.805} & \fancycellYellow{60.4s} & \fancycellYellow{\$0.07} & \fancycellGray{0.762} & \fancycellGray{89.3s} & \fancycellGray{\$0.09} & \fancyCellRed{n/a} & \fancyCellRed{n/a} & \fancyCellRed{n/a} & \fancycellGreen{0.820} & \fancycellGreen{14.3s} & \fancycellGreen{\$0.06} \\
    \textbf{Score} & \fancycellGray{0.576} & \fancycellGreen{17.9s} & \fancycellGreen{\$0.07} & \fancycellYellow{0.600} & \fancycellYellow{22.4s} & \fancycellGray{\$0.21} & \fancyCellRed{n/a} & \fancyCellRed{n/a} & \fancyCellRed{n/a} & \fancycellGreen{0.607} & \fancycellGray{22.7s} & \fancycellYellow{\$0.07} \\
    \textbf{Classify} & \fancycellYellow{0.763} & \fancycellGray{246.8s} & \fancycellYellow{\$1.08} & \fancycellGray{0.733} & \fancycellYellow{186.6s} & \fancycellGray{\$1.30} & \fancyCellRed{n/a} & \fancyCellRed{n/a} & \fancyCellRed{n/a} & \fancycellGreen{0.785} & \fancycellGreen{25.5s} & \fancycellGreen{\$0.73} \\
    \addlinespace[0.3ex]
    \midrule
    \addlinespace[0.3ex]
    \textbf{Avg (All Operators)} & \fancycellNormal{0.731} & \fancycellNormal{212.1s} & \fancycellNormal{\$0.86} & \fancycellNormal{0.715} & \fancycellNormal{165.0s} & \fancycellNormal{\$1.06} & \fancycellNormal{0.592} & \fancycellNormal{244.2s} & \fancycellNormal{\$0.17} & \fancycellNormal{0.677} & \fancycellNormal{31.7s} & \fancycellNormal{\$0.45} \\
    \textit{Avg (Filter + Join)} & \fancycellNormal{0.755} & \fancycellNormal{367.7s} & \fancycellNormal{\$1.54} & \fancycellNormal{0.740} & \fancycellNormal{263.2s} & \fancycellNormal{\$1.84} & \fancycellNormal{0.592} & \fancycellNormal{244.2s} & \fancycellNormal{\$0.17} & \fancycellNormal{0.587} & \fancycellNormal{48.1s} & \fancycellNormal{\$0.69} \\
    \bottomrule
  \end{tabular*}
  \endgroup
  \end{threeparttable}
\end{table*}

%% file: sections/07_conclusions.tex
\section{Conclusions and Outlook}
\label{sec:conclusions}

We introduced SemBench, a benchmark evaluating a novel class of systems: semantic query processing engines. Despite using only a small part of the available benchmark data for our experiments, we find that each of the evaluated systems incurs non-negligible processing overheads (cost and time) at least for some of the queries. Therefore, we believe that our benchmark will remain useful for stress-testing SQPEs in the coming years.

We identified several factors with a significant impact on the relative performance of SQPEs in our experiments. First, the implementation of semantic operators matters, for instance, for LIMIT and JOIN operators, where we observed significant differences in terms of costs as well as result quality. Second, prompt design matters for semantic operator implementations. Different SQPEs seem to aim at different cost-quality tradeoffs by using either relatively compact prompts, lowering costs, or more detailed prompts that tend to improve quality. 
An interesting extension to existing SQPEs would be automated prompt design strategies, optimizing cost-quality tradeoffs, or automatically rewriting prompts to fix situations in which the LLM refuses to generate answers.
At this point, none of the evaluated systems implement such strategies.

Several optimizations, currently not or not widely supported by SQPEs, seem useful to improve performance in our benchmark. For instance, fusing multiple semantic operators into one single LLM call offers potential for performance improvements in several scenarios (e.g., in Q10 and Q11 of the E-Commerce scenario). As our scenarios entail multiple queries on overlapping data subsets, some of them with equal or similar predicates, caching strategies would be useful to reduce processing overheads further. 

At present, no evaluated system supports all benchmark queries due to limitations in supported operators or data modalities. On the other hand, our benchmark includes scenarios in which almost all evaluated systems support all queries. This gives developers of SQPEs the opportunity to start evaluating early-stage versions of their systems, supporting a limited set of features, on easier scenarios, while expanding the scope to more SemBench scenarios as new features are added. As SQPEs expand the range of supported semantic operators over the coming years (e.g., semantic aggregation), we plan to add more scenarios in future SemBench instantiations.
